\documentclass[twoside,twocolumn,9pt]{article}
\usepackage{extsizes}
\usepackage[super,sort&compress,comma]{natbib} 
\usepackage[version=3]{mhchem}
\usepackage[left=1.5cm, right=1.5cm, top=1.785cm, bottom=2.0cm]{geometry}
\usepackage{balance}
\usepackage{packages/widetext}
\usepackage{times,mathptmx}
\usepackage{sectsty}
\usepackage{graphicx} 
\usepackage{lastpage}
\usepackage[format=plain,justification=raggedright,singlelinecheck=false,font={stretch=1.125,small,sf},labelfont=bf,labelsep=space]{caption}
\usepackage{float}
\usepackage{fancyhdr}
\usepackage{fnpos}
\usepackage[english]{babel}
\usepackage{array}
\usepackage{droidsans}
\usepackage{charter}
\usepackage[T1]{fontenc}
\usepackage[usenames,dvipsnames]{xcolor}
\usepackage{setspace}
\usepackage[compact]{titlesec}

\definecolor{cream}{RGB}{222,217,201}

\usepackage{textcomp}
\usepackage{upgreek}
\usepackage{color}

\begin{document}

\pagestyle{fancy}
\thispagestyle{plain}
\fancypagestyle{plain}{

\fancyhead[C]{\includegraphics[width=18.5cm]{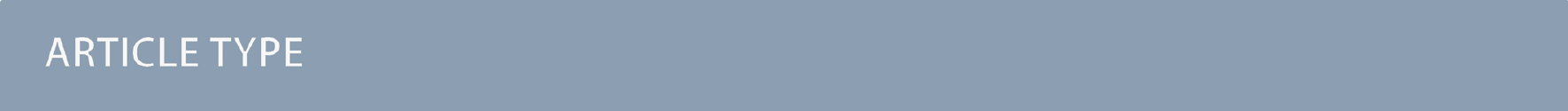}}
\fancyhead[L]{\hspace{0cm}\vspace{1.5cm}\includegraphics[height=30pt]{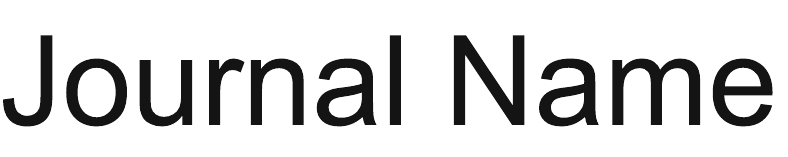}}
\fancyhead[R]{\hspace{0cm}\vspace{1.7cm}\includegraphics[height=55pt]{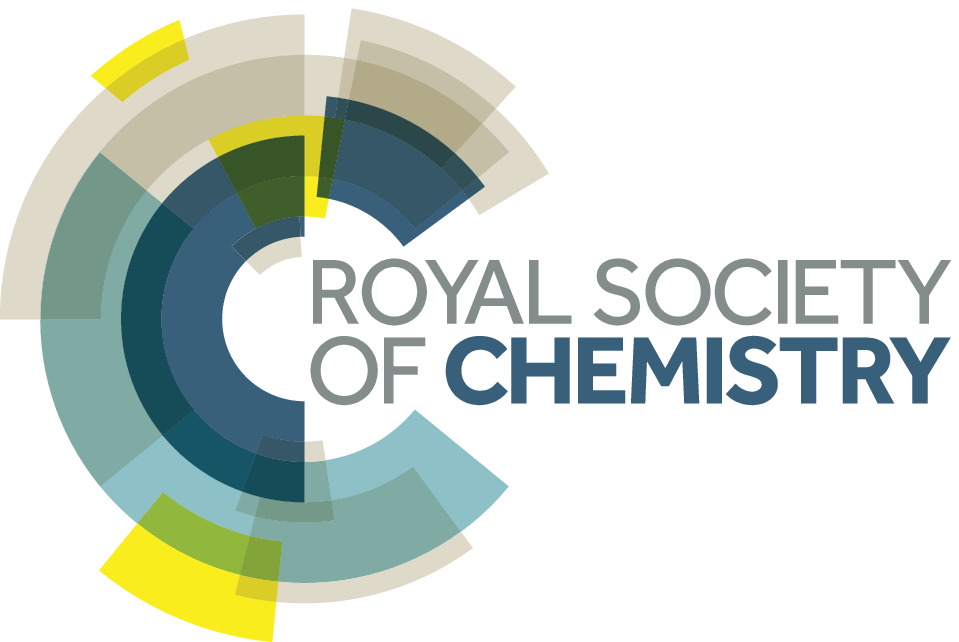}}
\renewcommand{\headrulewidth}{0pt}
}

\makeFNbottom
\makeatletter
\renewcommand\LARGE{\@setfontsize\LARGE{15pt}{17}}
\renewcommand\Large{\@setfontsize\Large{12pt}{14}}
\renewcommand\large{\@setfontsize\large{10pt}{12}}
\renewcommand\footnotesize{\@setfontsize\footnotesize{7pt}{10}}
\makeatother

\renewcommand{\thefootnote}{\fnsymbol{footnote}}
\renewcommand\footnoterule{\vspace*{1pt}%
\color{cream}\hrule width 3.5in height 0.4pt \color{black}\vspace*{5pt}} 
\setcounter{secnumdepth}{5}

\makeatletter 
\renewcommand\@biblabel[1]{#1}            
\renewcommand\@makefntext[1]
{\noindent\makebox[0pt][r]{\@thefnmark\,}#1}
\makeatother 
\renewcommand{\figurename}{\small{Fig.}~}
\sectionfont{\sffamily\Large}
\subsectionfont{\normalsize}
\subsubsectionfont{\bf}
\setstretch{1.125} 
\setlength{\skip\footins}{0.8cm}
\setlength{\footnotesep}{0.25cm}
\setlength{\jot}{10pt}
\titlespacing*{\section}{0pt}{4pt}{4pt}
\titlespacing*{\subsection}{0pt}{15pt}{1pt}

\fancyfoot{}
\fancyfoot[LO,RE]{\vspace{-7.1pt}\includegraphics[height=9pt]{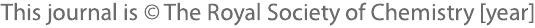}}
\fancyfoot[CO]{\vspace{-7.1pt}\hspace{13.2cm}\includegraphics{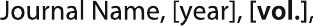}}
\fancyfoot[CE]{\vspace{-7.2pt}\hspace{-14.2cm}\includegraphics{head_foot/RF}}
\fancyfoot[RO]{\footnotesize{\sffamily{1--\pageref{LastPage} ~\textbar  \hspace{2pt}\thepage}}}
\fancyfoot[LE]{\footnotesize{\sffamily{\thepage~\textbar\hspace{3.45cm} 1--\pageref{LastPage}}}}
\fancyhead{}
\renewcommand{\headrulewidth}{0pt} 
\renewcommand{\footrulewidth}{0pt}
\setlength{\arrayrulewidth}{1pt}
\setlength{\columnsep}{6.5mm}
\setlength\bibsep{1pt}

\makeatletter 
\newlength{\figrulesep} 
\setlength{\figrulesep}{0.5\textfloatsep} 

\newcommand{\topfigrule}{\vspace*{-1pt}%
\noindent{\color{cream}\rule[-\figrulesep]{\columnwidth}{1.5pt}} }

\newcommand{\botfigrule}{\vspace*{-2pt}%
\noindent{\color{cream}\rule[\figrulesep]{\columnwidth}{1.5pt}} }

\newcommand{\dblfigrule}{\vspace*{-1pt}%
\noindent{\color{cream}\rule[-\figrulesep]{\textwidth}{1.5pt}} }

\makeatother

\twocolumn[
  \begin{@twocolumnfalse}
\vspace{3cm}
\sffamily
\begin{tabular}{m{4.5cm} p{13.5cm} }

\includegraphics{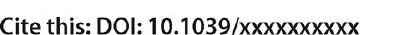} & \noindent\LARGE{\textbf{Microliter viscometry using a bright-field microscope: $\eta$-DDM}} \\
\vspace{0.3cm} & \vspace{0.3cm} \\

 & \noindent\large{M.~A. Escobedo-Sanchez,$^{\ast}$\textit{$^{a}$} J.~P. Segovia-Guti{\'e}rrez,\textit{$^{a,b}$} A.~B. Zuccolotto-Bernez,\textit{$^{c}$} J.~Hansen,\textit{$^{a}$}  C.~C.~Marciniak,\textit{$^{a}$} K.~Sachowsky,\textit{$^{a}$} F.~Platten,$^{\ast}$\textit{$^{a}$} and S.~U.~Egelhaaf\textit{$^{a}$}} \\
 
\includegraphics{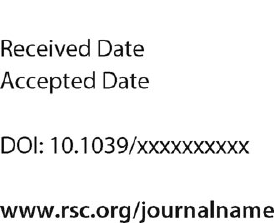} & \noindent\normalsize{Passive microrheology exploits the Brownian motion of colloidal tracer particles. From the mean-squared displacement (MSD) of the tracers, the bulk rheological and viscometric properties of the host medium can be inferred. Here, the MSD is determined by applying Differential Dynamic Microscopy (DDM). Compared to other microscopy techniques, DDM avoids particle tracking but provides parameters commonly acquired in light scattering experiments. Based on the spatial Fourier transform of image differences, the intermediate scattering function and subsequently the MSD is calculated. Then the usual microrheology procedure and the empirical Cox-Merz rule yield the steady-shear viscosity. This method, $\eta$-DDM, is tested and illustrated using three different systems: Newtonian fluids (glycerol-water mixtures), colloidal suspensions (protein samples) and a viscoelastic polymer solution (aqueous poly(ethylene oxide) solution). These tests show that common lab equipment, namely a bright-field optical microscope, can be used as a convenient and reliable microliter viscometer. Because $\eta$-DDM requires much smaller sample volumes than classical rheometry, only about a microliter, it is particularly useful for biological and soft matter systems.
} \\

\end{tabular}
\end{@twocolumnfalse} \vspace{0.6cm}
  ]

\renewcommand*\rmdefault{bch}\normalfont\upshape
\rmfamily
\section*{}
\vspace{-1cm}

\footnotetext{\textit{$^{a}$~Condensed Matter Physics Laboratory, Heinrich Heine University, 40225 D\"usseldorf, Germany. E-mail: escobedo@hhu.de, Florian.Platten@hhu.de}}
\footnotetext{\textit{$^{b}$~Department of Applied Physics, Faculty of Sciences, University of Granada, Fuentenueva s/n, 18071 Granada, Spain.}}
\footnotetext{\textit{$^{c}$~Departamento de F{\'i}sica, CINVESTAV-IPN, Av.~Instituto Polit{\'e}cnico Nacional 2508, 07360 M{\'e}xico D.F, Mexico.}}




\section{Introduction}
	\label{intro}

One of the main rheological parameters is the shear viscosity $\eta$ which quantifies the resistance of a material against flow.\cite{Walters1975, Macosko1994, Larson1999} Apart from the fundamental-physics interest in the shear viscosity, the determination of its magnitude is a common way to characterize, for instance, macromolecular solutions.\cite{Larson1999, Tanford1963} The viscosity of dilute solutions depends on the size, shape and mass of the macromolecules. In concentrated solutions, the viscosity also reflects the interactions between the macromolecules and can indicate, e.g., structural changes or aggregation.\cite{Nicoud2015, Lindsay1982, Schoepe1998, Zhou2015}

Conventional viscometry, e.g.~using a falling ball viscometer or Ostwald viscometer tube, has successfully been used to measure shear viscosity.\cite{Goodwin2014} 
However, these methods require relatively large sample volumes, typically on the order of milliliters. This is a significant disadvantage for expensive or  scarce samples. The development of new capillary viscometers has reduced this issue because it only needs a small sample volume, on the order of microliters.\cite{Hudson2015} 
In addition, modern torsional rheometers  
also require only small sample volumes and offer the possibility to finely control either the applied shear stress or shear strain. Rotational rheometers allow for further stress conditions: creep flows, start-up tests, squeeze tests, oscillatory shear tests etc.\cite{Macosko1994, Larson1999} For instance, oscillatory shear measurements provide important information about the system's response to external stimuli within a wide range of frequencies. This technique allows one to obtain frequency-dependent parameters, in particular the viscoelastic complex modulus $G^*(\omega)=G'(\omega)'+iG''(\omega)$. It displays the system's ability to store ($G'$) or dissipate ($G''$) energy and is related to the complex viscosity $\eta^*(\omega)=G^*(\omega)/i\omega$.
An accurate estimate of the steady-shear viscosity can be obtained applying oscillatory tests together with the empirical Cox-Merz rule.\cite{CoxMerz1958} This rule relates the linear viscoelastic response determined in oscillatory measurements to the nonlinear behaviour measured in steady shear flow tests. It states that
\begin{equation}
\label{coxmerz}
    |\eta^*(\omega)|_{\omega\rightarrow{0}} \equiv \eta(\dot{\gamma})_{\dot{\gamma}\rightarrow{0}} \;\; ,
\end{equation}
where $ |\eta^*(\omega)|$ is the modulus of the complex viscosity and $\dot{\gamma}$ is the shear rate. The applicability of the Cox-Merz rule is restricted to low frequencies and low shear rates and thus to small amplitude oscillatory shear (SAOS) measurements, in which a frequency sweep is performed with a constant small strain amplitude within the linear viscoelastic region. Its validity has been successfully examined for a wide variety of complex systems, especially for polymeric solutions.\cite{McKinley2012} Hence SAOS measurements can replace conventional viscometry measurements when reliable information cannot be obtained from steady-shear measurements due to the limited torque resolution.

Passive microrheology\cite{Mason1995, Waigh2005, Cicuta2007} can be considered the counterpart of SAOS measurements on the microscale. It exploits the thermal motion of tracer particles which undergo only small displacements due to their low (thermal) energy. Thus, passive microrheology provides information on the viscous and elastic properties within the linear viscoelastic region, similar to SAOS experiments. Nevertheless, it extends the frequency range to values that, due to the torque resolution limit, are inaccessible to conventional torsional rheometers. Furthermore, the possibility to use even smaller sample volumes, only a few microliters, renders this technique a very convenient tool for precious samples of which only small volumes are available.

Passive microrheology is based on the Brownian motion of colloidal tracer particles.\cite{Waigh2005, Cicuta2007} Care has to be taken that the tracers are stable and do not significantly alter the system, e.g., through interactions with or adsorption to the tracer surface,\cite{Valentine2004, Qiu09, Dasgupta2002} but also that the tracers comprehensively sample the system, e.g., in the presence of heterogeneities in particle density or polymer network structure.\cite{Gardel05, Waigh2005}
The dynamics of the tracers, namely their mean-squared displacement (MSD), is related to the rheological and viscometric properties of the solution through the generalized Stokes-Einstein relation.\cite{Squires2010} The MSD of the tracers can be extracted by, e.g., multiple particle tracking (MPT).\cite{Crocker1996, Mason1997} This requires to determine the positions of the tracers using time series of micrographs and to link them to trajectories.\cite{Crocker1996} This procedure is prone to artifacts and hence needs a careful selection of trajectories.\cite{Kowalczyk2015} Despite these limitations, MPT has successfully been used to determine, for instance, the viscosity of protein solutions\cite{Tu2005, Josephson2016} as well as the mechanical properties of biological and soft matter systems.\cite{Valentine2001, Chen2010} Moreover, the MSD and hence the rheological properties can also be determined using light scattering techniques, both dynamic light scattering (DLS) and diffusing wave spectroscopy (DWS) have been applied.\cite{Mason1995, Dasgupta2002} The light scattering methods, in particular DWS, typically probe much larger sample volumes than microscopy techniques and hence provide superior statistics or shorter measurement times. Furthermore, the temporal and spatial resolution, again in particular of DWS, is significantly improved compared to microscopy techniques and therefore allows for the investigation of very high frequencies, up to about $10^7$~rad/s, and of highly viscous samples, in which the tracer motion is severely limited.

Another, more recent family of techniques to study the dynamics of Brownian particles is Digital Fourier Microscopy (DFM).\cite{Giavazzi2009, Giavazzi2014, Cerbino2008, Cerbino2017} It uses different imaging techniques to obtain spatio-temporal information on biological, soft matter and other systems.\cite{Wilson2011, Martinez2012, Germain2016, Bayles2016} If the images are acquired using bright field microscopy, the technique is also known as Differential Dynamic Microscopy (DDM).\cite{Cerbino2008, Cerbino2017} A spatial Fourier analysis of image differences provides quantities that are identical to the ones extracted from light scattering experiments.\cite{Giavazzi2014} In particular, it provides access to the intermediate scattering function $f({\textbf{Q}},\tau)$ with the scattering vector $\textbf{Q}$ and the delay time $\tau$, which is commonly acquired in DLS and DWS experiments. Based on $f({\textbf{Q}},\tau)$, the MSD of the tracers can be inferred.
Very recently, DDM has been applied to passive microrheology.\cite{Bayles2017, Edera2017}

The analogy between passive microrheology and SAOS together with the Cox-Merz rule suggest that the steady-shear viscosity $\eta$ can be determined using DDM, a method we will call $\eta$-DDM. We test $\eta$-DDM and its applicability for three different systems of increasing complexity: Newtonian fluids (glycerol-water mixtures), colloidal samples (aqueous solutions containing a globular protein) and a viscoelastic system (a polymer solution containing poly(ethylene oxide)). For all systems, the results obtained by $\eta$-DDM quantitatively agree with independent glass capillary, rotational rheometer and dynamic light scattering measurements as well as literature data. This also confirms the applicability of the Cox-Merz rule. Although $\eta$-DDM relies on real-space images, it does not involve particle tracking and thus avoids the limitations of MPT experiments. Furthermore, it does not need special equipment, such as capillary viscometers.\cite{Grupi2012, Hudson2015} A conventional bright-field microscope can be used as a convenient and reliable viscometer that only requires very small sample volumes, about a microliter.

The remainder of the manuscript is structured as follows: the method to determine the steady-shear viscosity $\eta$ based on DDM data is presented in Section~\ref{sec:DDM}, the sample preparation procedures and applied techniques are described in Section~\ref{sec:mat}, while in Section~\ref{sec:results} the proposed method, $\eta$-DDM, is tested using three different systems and, finally, the characteristics of $\eta$-DDM are summarized in Section~\ref{sec:conclusions}.


\bigskip
\section{Determination of the Steady-Shear Viscosity in a DDM Experiment}
\label{sec:DDM}

DDM is used to determine the tracer dynamics. In a DDM experiment, a time series of images $i(\textbf{x},t)$ is taken and their differences computed according to
\begin{equation} \label{eq:idiff}
    \triangle i(\textbf{x},t,\tau)=i(\textbf{x},t{+}\tau)-i(\textbf{x},t)
\end{equation} 
with $\textbf{x}$ the position in the imaged plane and $\tau$ the delay time between two images. The subsequent analysis yields the structure function in reciprocal space which is defined as\cite{Giavazzi2009, Giavazzi2014, Cerbino2008, Cerbino2017}
\begin{equation}\label{timeaveragedD}
    D(Q,\tau)=\langle |\mathcal{F}\left[ \triangle i(\textbf{x},t,\tau) \right]|^{2} \rangle_{t,\phi}  \;\;  ,
\end{equation}
where the modulus of the scattering vector $Q = (4\pi n/\lambda_0) \sin{(\theta/2)}$ with $n$ the refractive index in the medium of propagation, $\theta$ the scattering angle and $\lambda_0$ the wavelength of light in vacuum. Moreover, ${\mathcal{F}}$ denotes a Fourier transform. While performing the numerical Fourier transform, care has to be taken to avoid imaged inhomogeneities leading to spectral leakage.\cite{Giavazzi2017_W} Assuming a stationary and isotropic signal, the signal statistics and hence the power spectrum are independent of time $t$ and the direction of the scattering vector $\textbf{Q}$. To improve statistics, thus, a temporal average over $t$ and an azimuthal average over the angle $\phi$ in the $\textbf{Q}$ plane are performed and indicated by $\langle \;\; \rangle_{t,\phi}$.

The structure function $D(Q,\tau)$ is related to the intermediate scattering function $f(Q,\tau)$ by\cite{Giavazzi2009, Giavazzi2014, Cerbino2008, Cerbino2017}
\begin{equation}\label{structure}
    D(Q,\tau)=A(Q)\left[1-f(Q,\tau) \right]+B(Q)  \;\; ,
\end{equation}
where the first term, $A(Q)$, contains information on the particles and $B(Q)$ is the power spectrum of the camera noise that unavoidably is present even in the absence of particles.\cite{Giavazzi2017_W} Both, $A(Q)$ and $B(Q)$ are estimated following a similar framework as previously proposed.\cite{Bayles2017}
The intermediate scattering function $f(Q,\tau)$ can be related to the MSD, $\langle \triangle {\textbf{r}}^2(\tau) \rangle$, through\cite{Cerbino2017, Megen89}
\begin{equation}\label{ISF1}
    f(Q,\tau)=\exp\left[-\frac{Q^2}{2d}\langle \triangle {\textbf{r}}^2(\tau)\rangle \right]  \;\; .
\end{equation}
The spatial dimension $d=2$, since a projection of the particle motion onto a two-dimensional detector is recorded. Combining Eqs.~\ref{structure} and \ref{ISF1}, the MSD of the tracers is obtained via the relation\cite{Bayles2017, Edera2017}
\begin{equation}\label{ISF2}
    \langle \triangle {\textbf{r}}^2(\tau)\rangle =-\frac{2d}{Q^2}\ln\left[1-\frac{D(Q,\tau)-B(Q)}{A(Q)} \right]  \;\; ,
\end{equation}
Therefore, DDM provides the MSD via the intermediate scattering function, which is commonly acquired in light scattering experiments. Particle tracking thus is not required. Furthermore, the tracers need not to be resolved and hence also small tracers can be used.\cite{Edera2017}

The MSD can be related to the rheological properties and the viscosity of the medium, especially the complex modulus $G^*(\omega)$, via the generalized Stoke-Einstein relation:\cite{Dasgupta2002, Mason1995}
\begin{equation}\label{Gstar1}
    G^{*}(\omega)=\frac{2d \, k_{B}T}{6\pi \, a \,  i \omega \, \mathcal{F}\{ \left< \triangle {\textbf{r}}^2(\tau)\right> \} } \;\; ,
\end{equation}
where $k_{B}$ is the Boltzmann constant, $T$ the temperature and $a$ the tracer radius. Assuming a local power law for $\langle \triangle {\textbf{r}}^2(\tau) \rangle$, one can define $\alpha(\omega) = \left | \partial \ln \langle \triangle {\textbf{r}}^2(\tau)\rangle /\partial \ln\tau \right |_{\tau=1/\omega}$ which accounts for the characteristics of the sample and takes values between $0$ and $1$.\cite{Mason2000, Dasgupta2002} If viscous behaviour dominates, the tracers show diffusive dynamics and $\alpha \approx 1$, whereas if elastic behaviour dominates, then $\alpha \approx 0$. With this assumption, the magnitude of the complex modulus $G^{*}(\omega)$ becomes
\begin{equation}\label{Gstar2}
    G(\omega)=\frac{2d \, k_{B}T}{6\pi \, a \,  \langle \triangle {\textbf{r}}^2(1/\omega)\rangle \, \Gamma\left[1{+}\alpha(\omega) \right]  }
\end{equation}
where $\Gamma$ denotes the gamma function.
Typically, one considers the real and imaginary parts of $G^{*}(\omega)$, i.e.~the elastic, $G'(\omega)$, and loss, $G''(\omega)$, moduli, respectively,
\begin{equation} \label{moduli}
\begin{split}
G'(\omega) & = G(\omega)\cos\left[\pi\alpha(\omega)/2 \right]  \;\;  ,   \\
G''(\omega) & = G(\omega)\sin\left[ \pi\alpha(\omega)/2\right]  \;\;  .
\end{split}
\end{equation}

Finally, the complex viscosity $\eta^*(\omega)$ can be calculated using its definition
\begin{equation}\label{EtaStarEq}
    \eta^*(\omega) = \frac{G^*(\omega)}{i\omega} \, .
\end{equation}
Exploiting the Cox-Merz rule (Eq.~\ref{coxmerz}), the corresponding steady-shear viscosity $\eta(\dot{\gamma})_{\dot{\gamma}\rightarrow{0}}$ can be estimated.


\bigskip
\section{Materials \& Methods}
\label{sec:mat}

\subsection{Sample preparation}

Glycerol (analytical reagent grade; Fisher Chemical, prod.~no.~G/0650/15), poly(ethylene oxide) (PEO, 900 kDa; Sigma-Aldrich, prod.~no.~189456),  lyophilized hen egg-white lysozyme (Roche Diagnostics, prod.~no.~10837059001), sodium acetate (NaAc, p.a.; Merck, prod.~no.~1.06268), sodium chloride (NaCl; Fisher Chemical, prod.~no.~148717) and choloroform (99.8~\%; Acros Organics, prod.~no.~404635000) were used without further purification.
Ultrapure water with a minimum resistivity of 18~M$\Omega$cm was used (Purelab$^{\text \textregistered}$ flex, Elga).

Polystyrene spheres (diameter 330~nm; Invitrogen) were used as tracers for glycerol-water mixtures and the PEO-water solution. For protein solutions, polystyrene spheres coated with hydrophilic PEG~300 (diameter~$1\;\upmu$m; micromer$^{\text \textregistered}$ particles, micromod Partikeltechnologie GmbH) were used as tracers that minimize protein-particle interactions.\cite{Valentine2004}  
Stock suspensions of uncoated and coated polystyrene spheres in ultrapure water were prepared with volume fractions of $0.076$ and $0.001$, respectively. The tracer volume fractions in the different samples are given below and were optimized to obtain good statistics while excluding tracer-tracer interactions.

All measurements were conducted at $T = 20^\circ$~C.

\subsubsection{Glycerol-water mixtures}

The purity of glycerol was determined by measuring its shear viscosity, $\eta \approx 1.218$~Pa\,s. This suggests a water mass fraction of about $0.68~\%$ when compared to literature values.\cite{Cheng2008} This water content was taken into account when preparing the glycerol-water mixtures. Glycerol-water mixtures were prepared by mixing appropriate amounts of water and glycerol to yield six different mass fractions of glycerol ranging from $0~\%$ to $56.8~\%$.
For the DDM measurements, small amounts of tracer stock suspensions were added to the glycerol-water mixtures to yield a tracer volume fraction of $3.75 \times 10^{-5}$.

\subsubsection{Lysozyme samples}

The protein powder was dissolved in a $50$~mM NaAc buffer solution which was adjusted to $p$H~$4.5$ by adding small amounts of hydrochloric acid. At this $p$H, lysozyme carries $11.4$ net positive elementary charges.\cite{Tanford1972} This solution condition closely resembles those used in previous studies.\cite{Sedgwick2005, Pan2009, Hansen2015b, Platten2015c, Godfrin2015, Platten2015} A protein solution with an initial protein concentration $c_\text{p} \approx 40$~mg/ml was filtered several times using a syringe filter with low protein binding (pore size $0.1~\upmu$m; Pall, Acrodisc, prod.~no.~4611) in order to remove impurities and undissolved proteins. Subsequently, it was concentrated by a factor of four to seven using a centrifugal filter (Amicon Ultra-15, PLGC Ultracel-PL Membran, $10$~kDa, Merck, prod.~no.~UFC901008) or a stirred ultra-filtration cell (Amicon, Millipore, prod.~no.~5121) with an Omega membrane disc filter - 10K (Pall, prod.~no.~OM010025). The retentate was used as protein stock solution. Its concentration was determined using a refractometer\cite{Platten2015b} and the protein volume fraction $\phi = c_\text{p} v_\text{p}$ calculated using the specific volume of lysozyme $v_\text{p} = 0.740~\text{cm}^3/\text{g}$.\cite{Platten2015b} Proper amounts of buffer, protein stock solution and tracer stock suspension were mixed to obtain the desired volume fractions of protein and tracers. The samples for the DDM measurements contained tracers with a volume fraction of about $5\times10^{-5}$.

\subsubsection{Aqueous poly(ethylene oxide) solution}

PEO powder was dissolved in ultrapure water while gently stirring at a constant temperature $T=40^\circ$~C until no undissolved polymer could visually be observed. To prevent bacteria proliferation, a few drops of chloroform ($\sim 0.1$~ml) were added to the PEO solution ($\sim 20$~ml). The addition of chloroform led to the formation of PEO aggregates that did not easily redissolve. Thus samples were kept at $40^\circ$~C for $10$~days during which undissolved polymer settled. Clean supernatant was carefully removed and placed in a new glass flask. The solution contained a PEO mass fraction of $2.1$~$\%$.
A small amount of tracer stock suspension was added to the PEO solution to yield a tracer volume fraction of $7.5\times10^{-4}$ (DDM) and $3.125\times10^{-4}$ (DLS), respectively.

\subsection{Differential Dynamic Microscopy (DDM)}
\label{subsec:DDM}

DDM experiments were performed with two different microscopes, an inverted (Nikon Ti-Eclipse) and an upright (Nikon Eclipse 80i) microscope, both equipped with a $20 \times$ microscope objective with a numerical aperture ${\text{NA}} = 0.5$. The power of the microscope lamp was tuned to adapt the illumination to the different samples. Images were acquired with an $8$~bit CMOS black and white camera with $1280\times 1024$~pixels, each with an area of $4.8\times 4.8\,\upmu$m$^2$ (Mako-U130, Allied Vision Technologies). 
The field of view was carefully selected to avoid impurities such as dust, tracer aggregates or undissolved small polymer lumps.
Measurement configurations were adapted to the characteristics of the samples: Series of 20,000 (glycerol-water), 50,000 (lysozyme) or 125,000 (PEO) images with $512 \times 512$ (glycerol-water, lysozyme) or $256 \times 256$ (PEO) pixels were recorded with a rate of 100 (glycerol-water, PEO) or 50 (lysozyme) frames per second with an exposure time of $1$~ms (glycerol-water, lysozyme) or $0.5$~ms (PEO). They were analyzed averaging 3,072 (glycerol-water, lysozyme) or 10,000 (PEO) image pairs per delay time (Eq.~\ref{eq:idiff}). Data were extracted for a $Q$ range of $[0.7, 4.75] \; \upmu \text{m}^{-1}$ (glycerol-water), $[0.7, 3.5] \; \upmu \text{m}^{-1}$ (lysozyme) or $[1.6, 4.75] \; \upmu \text{m}^{-1}$ (PEO).

The glycerol-water mixtures and the lysozyme solutions were imaged in home-built sample cells consisting of a microscope slide and three cover slips glued together to form a small capillary.\cite{Jenkins2008} The PEO solutions were kept in commercial capillaries with a rectangular cross-section (inner dimensions $0.20 \times 2.00$~mm, Vitrotubes$^{\text \textregistered}$, VitroCom, CM Scientific).
In the case of the viscoelastic PEO sample, the rheological properties exhibited a time dependence after loading the sample into the capillary. It disappeared after long enough equilibration. Although this is an interesting observation, it is beyond the scope of the current work. Thus, before performing DDM measurements with the PEO sample, it was left to equilibrate for three days.

\subsection{Dynamic light scattering (DLS)}

Homodyne far-field light scattering experiments were carried out using a 3D dynamic light scattering apparatus (LS Instruments AG) with a He-Ne laser (wavelength $\lambda_{0} = 632.8$~nm, power $32$~mW, JDSU), a pair of avalanche photodiodes (Perkin-Elmer) and a multitau digital correlator. Standard DLS (not 3D-DLS) measurements were performed at three scattering angles, $\theta = 20.4^\circ$, $22.2^\circ$ and $24.2^\circ$, which correspond to $Q = 4.69 \; \upmu \text{m}^{-1}$, $5.09 \; \upmu \text{m}^{-1}$ and $5.54 \; \upmu \text{m}^{-1}$, respectively. This range overlaps with the $Q$ range of the DDM experiments. The sample was held in a cylindrical cuvette with an inner diameter of $8.5$~mm.

Based on the measured intensity trace, $I(t)$, the correlator provides the normalized time-averaged autocorrelation function of the scattered intensity\cite{Pecora}
\begin{equation}\label{G2}
      g^{(2)}(Q,\tau)= \frac{\langle I(Q,t) \, I(Q,t{+}\tau) \rangle_t}{\langle I(Q,t) \rangle_t^{2}} \;\;  ,
\end{equation}
where $\langle \;\; \rangle_t$ represents a time-average. The $g^{(2)}(Q,\tau)$ can be related to the intermediate scattering function $f(Q,\tau)$ through the Siegert relation\cite{Pecora}
\begin{equation}\label{Siegert}
      g^{(2)}(Q,\tau)=1+\beta|f(Q,\tau)|^{2}   \;\;  ,
\end{equation}
where $\beta$ is the intercept. Again, $f(Q,\tau)$ is related to the MSD of the tracers (Eq.~\ref{ISF1}) but now $d=3$. This allows to calculate the complex shear modulus and the steady-shear viscosity (Eqs.~\ref{Gstar1} -- \ref{EtaStarEq}).

\subsection{Macroscopic rheology and viscometry}

Rheological tests were carried out using a strain-controlled rheometer (ARES G2, TA Instruments) equipped with a cone with a diameter of $50$~mm, an angle of $0.02$~rad and a truncation gap of $51~\upmu$m. Samples with volumes of about $0.7$~ml  were placed onto the fixed bottom plate of the geometry avoiding any air bubbles and enclosed by a solvent trap containing pads soaked with water to reduce solvent evaporation. 
To remove effects of the filling procedure and improve reproducibility, after filling a two-step procedure was performed: a flow ramp from $\dot{\gamma} = 10~\text{s}^{-1}$ to $0~\text{s}^{-1}$ during a total time of $90$~s followed by a relaxation at $0~\text{s}^{-1}$ for $30$~s.

Small-amplitude oscillatory shear (SAOS) tests were performed in the linear viscoelastic region with a strain amplitude $\gamma_0 = 0.09$ and a frequency interval $100~\text{rad} \; \text{s}^{-1} \le \omega \le 0.1~\text{rad} \; \text{s}^{-1}$. The amplitude value was selected well within the linear viscoelastic regime, determined by performing an amplitude sweep ranging from $\gamma_0 = 0.005$ to $2$ at a constant angular frequency $\omega = 1~\text{rad}/\text{s}$.  
The measurement time per frequency was automatically selected by the rheometer software for each frequency.

Steady-shear measurements (flow curves) were performed from $\dot{\gamma} = 10~\text{s}^{-1}$ to $1000~\text{s}^{-1}$ with an individual measurement time $t_{\text{m}}=12$~s in the case of the glycerol-water mixtures and from $\dot{\gamma} = 0.1~\text{s}^{-1}$ to $100~\text{s}^{-1}$ with $t_{\text{m}}=100$~s (to achieve an appropriate sampling in the small shear rate regime) in the case of the aqueous PEO solution.

To improve statistics, in all tests three identical samples were measured each three times.

Capillary viscometry was conducted using an Ubbelohde glass capillary viscometer (Schott) which was thoroughly cleaned before use. Each measurement typically required a volume of about $12$~ml.


\bigskip
\section{Results and Discussion}
\label{sec:results}

The viscosities of different model systems were determined using DDM and the procedure described in Sec.~\ref{sec:DDM}. The measurements are based on the determination of the tracer dynamics, namely the intermediate scattering function $f(Q,\tau)$ and the MSD $\langle \triangle {\textbf{r}}^2(\tau) \rangle$, from which the rheological properties, in particular the steady-shear viscosity $\eta(\dot{\gamma})_{\dot{\gamma}\rightarrow{0}}$, are calculated and compared to results from other techniques. This allows us to test the validity of $\eta$-DDM through a comparison with classical rheology measurements und dynamic light scattering experiments. This is illustrated for three model systems of increasing complexity: first, Newtonian fluids with different viscosities realized by mixtures of two simple liquids, namely water and glycerol. Second, a particle suspension with some biological relevance, i.e.~a protein (lysozyme) in buffer with different protein concentrations. Third, a more complex situation with a viscoelastic fluid represented by a polymer solution containing poly(ethylene oxide) in water.


\subsection{Newtonian fluids: glycerol-water mixtures}

$\eta$-DDM is applied to mixtures of simple liquids with different viscosities, namely glycerol-water mixtures with glycerol mass fractions ranging from $0\;\%$ to about $57\;\%$. Glycerol-water mixtures have frequently been examined,\cite{Glycerine1, Glycerine2, Glycerine3} including their viscosity.\cite{Segur1951, Gonzalez2011} An empirical formula for their viscosity over a wide temperature and concentration range has been established.\cite{Cheng2008}  As a consequence, glycerol-water mixtures have become a reference system to test viscometry and microrheology techniques.\cite{Edera2017}

\begin{figure}[bt!]
\begin{centering}
\includegraphics[width=0.76\linewidth]{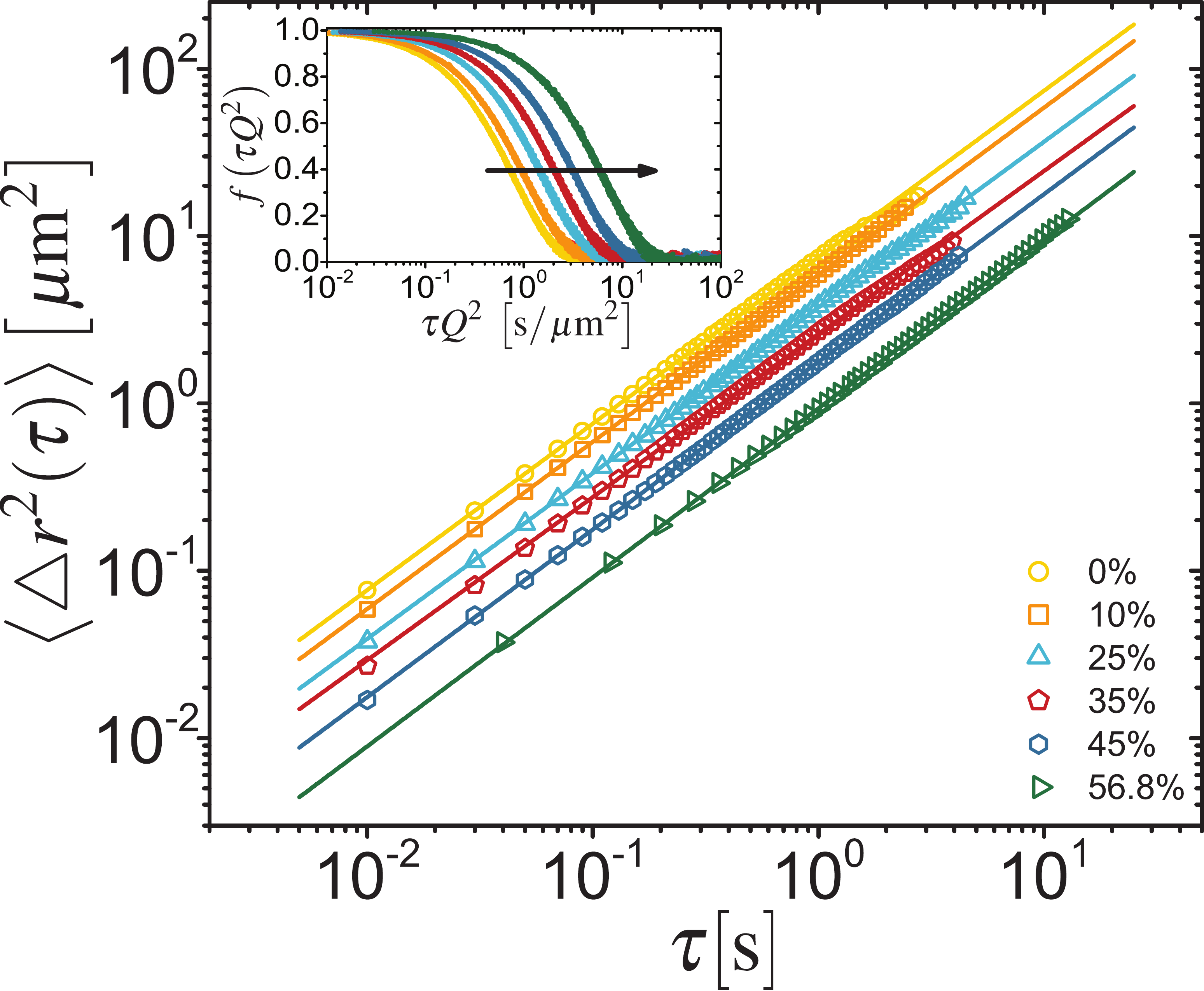}
\par\end{centering}
\protect\caption{Mean-squared displacement (MSD), $\langle \triangle r^2(\tau) \rangle$, as a function of delay time $\tau$ as determined by $\eta$-DDM for tracers with a diameter of $330$~nm in glycerol-water mixtures with different glycerol mass fractions (as indicated). Inset: Corresponding intermediate scattering function $f(Q,\tau)$ as a function of $\tau Q^2$. Note that, in this representation, $f(Q,\tau)$ for different $Q$ fall on top of each other. The arrow indicates increasing glycerol mass fractions.}
\label{fQtMSDWG}
\end{figure}

\begin{figure}[bt!]
\begin{centering}
\includegraphics[width=0.76\linewidth]{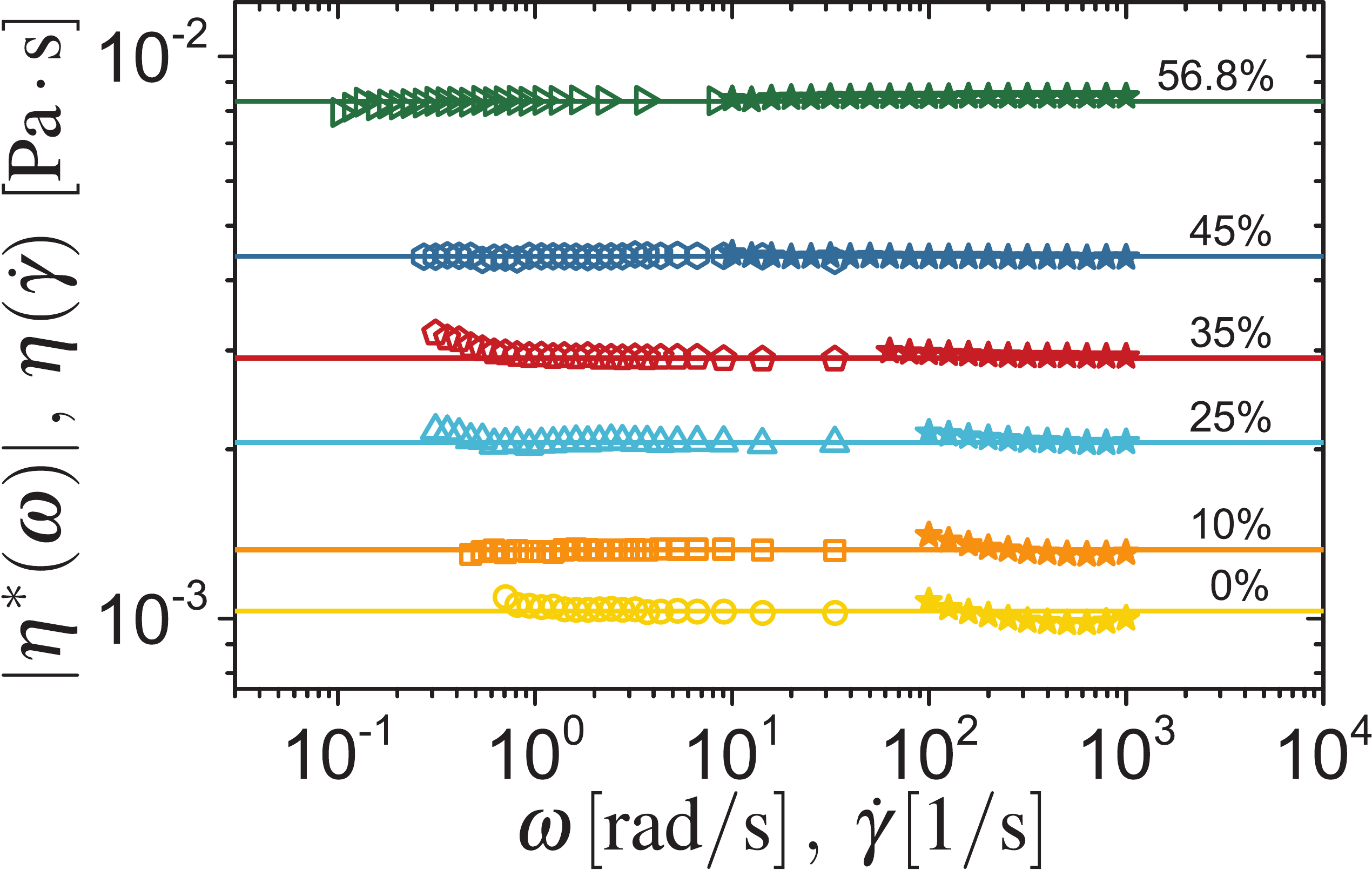}
\par\end{centering}
\protect\caption{Modulus of the complex viscosity $| \eta^{*}(\omega) |$ and steady-shear viscosity $\eta(\dot{\gamma})$ as a function of angular frequency $\omega$ and shear rate $\dot{\gamma}$ as determined by $\eta$-DDM (open symbols) and steady-shear measurements (filled stars), respectively, of glycerol-water mixtures with different glycerol mass fractions (as indicated).}
\label{complexWG}
\end{figure}

\begin{figure}[bt!]
\begin{centering}
\includegraphics[width=0.76\linewidth]{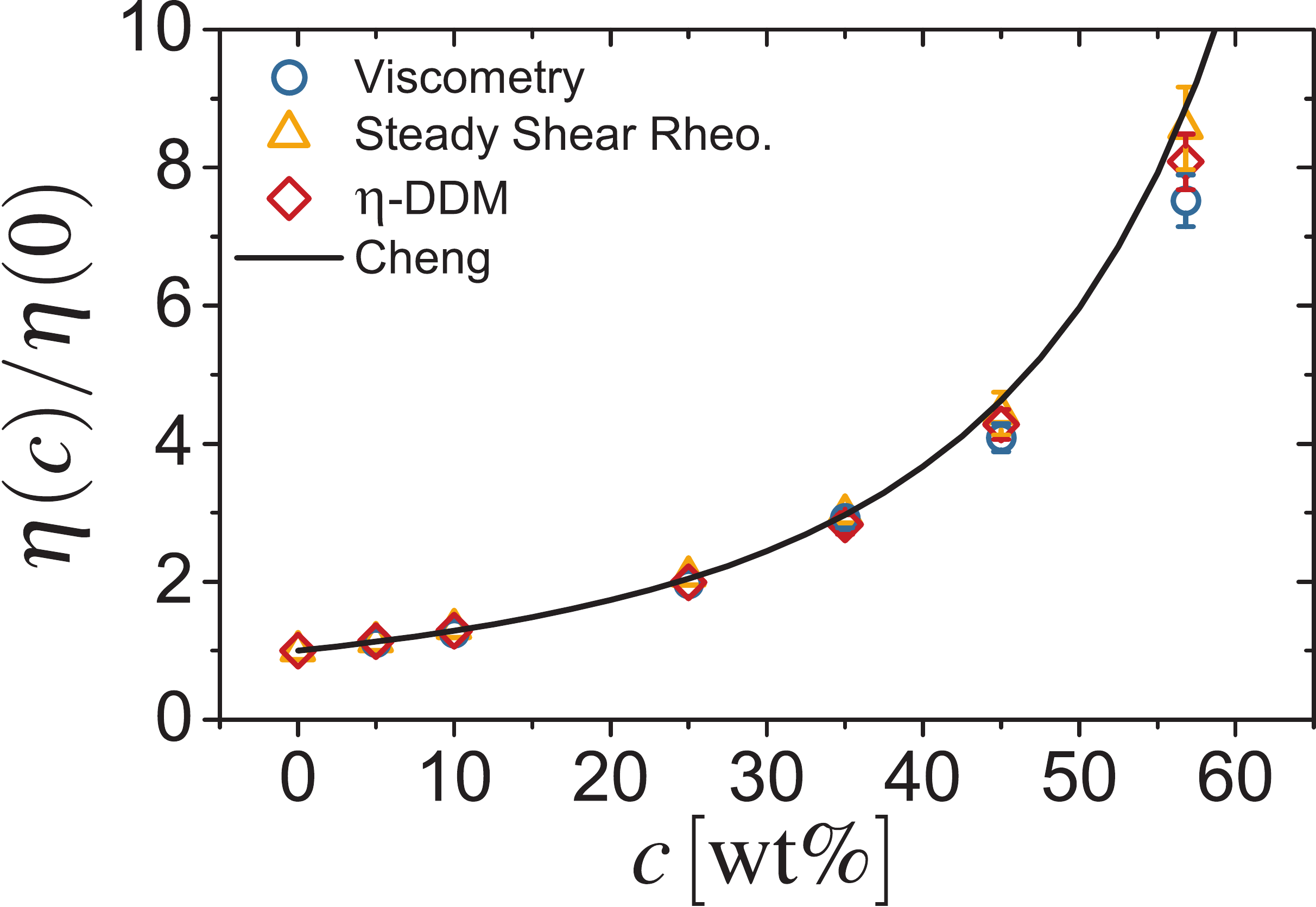}
\par\end{centering}
\protect\caption{Normalized viscosity, $\eta(c)/\eta(0)$, as a function of glycerol mass fraction $c$ of glycerol-water mixtures as determined by $\eta$-DDM, steady-shear and viscometry measurements as well as literature data\cite{Cheng2008} (as indicated).}
\label{AllW}
\end{figure}

In our DDM experiments, series of images were recorded and analyzed via the structure function $D(Q,\tau)$ (Sec.~\ref{sec:DDM}). This analysis yields the intermediate scattering function $f(Q,\tau)$. If represented as a function of $\tau Q^2$, the $f(Q,\tau)$ obtained for a specific glycerol content but different $Q$ values fall on top of each other (Fig.~\ref{fQtMSDWG}, inset). This in particular implies that the tracers are not affecting each other. Thus, the tracer concentration was low enough and the tracer condition fulfilled.\cite{Megen89} With increasing glycerol content, the decay of $f(Q,\tau)$ occurs at later delay times $\tau$. This reflects the slower diffusion of the tracers due to the increasing viscosity of the liquid mixture.
In the further analysis, only data points with a sufficient signal-to-noise ratio are considered, thus delay times $\tau$ were restricted to $f(Q,\tau) > 0.1$. This implies that the range of $\tau$ depends on the mixture, namely its viscosity. (The range of $\tau$ also depends on the size of the tracers, which hence can be used to tune the delay-time window.)
The MSDs for the different glycerol-water mixtures were extracted (Eq.~\ref{ISF2}, Fig.~\ref{fQtMSDWG}).
The shift in the MSDs also reflects the slower diffusion upon increasing the glycerol content. Within the accessible time window, the MSDs show a linear dependence on the delay time $\tau$. This implies a purely diffusive motion and is consistent with the Newtonian (viscous) nature of the system. 
Subsequently, the modulus of the complex viscosity $|\eta^*(\omega)|$ was calculated as a function of the angular frequency $\omega$ (Fig. \ref{complexWG}). Consistent with the Newtonian behaviour of the system, $|\eta^*(\omega)|$ does not depend on the frequency $\omega$ and hence an extrapolation to $\omega \rightarrow 0$ is unambiguously possible. Application of the Cox-Merz rule (Eq.~\ref{coxmerz}) yields the steady-state viscosity (Fig.~\ref{AllW}). It shows the expected increase with glycerol concentration.

For comparison, steady-shear measurements were performed to determine the viscosity $\eta(\dot{\gamma})$ as a function of the shear rate $\dot{\gamma}$ (Fig.~\ref{complexWG}, filled stars). No significant dependence on $\dot{\gamma}$ is observed. The steady-shear viscosity $\eta(\dot{\gamma})$ and the magnitude of the complex viscosity $|\eta^*(\omega)|$, including their (in)dependence on $\dot{\gamma}$ and $\omega$, agree for all studied glycerol concentrations. This confirms the Cox-Merz rule (Eq.~\ref{coxmerz}) and the validity of the $\eta$-DDM measurements. Only for the two highest glycerol concentrations and hence the two highest viscosities, the frequency windows of the steady-shear and DDM measurements overlap (Fig.~\ref{complexWG}). This is due to the limited sensitivity of the rheometer which, for the other glycerol concentrations, does not allow for measurements below about $100 \; \text{s}^{-1}$. Thus, for systems with a very low viscosity, $\eta$-DDM complements mechanical rheometry at low frequencies. Finally, the agreement between $\eta$-DDM and  steady-shear measurements is also reflected in the identical dependence of the viscosity on the glycerol concentration (Fig.~\ref{AllW}).

The viscosity $\eta$ of the glycerol-water mixtures was also investigated by viscometry. The determined viscosities are normalized by the one measured for pure water (Fig.~\ref{AllW}). The obtained dependence on the glycerol concentration $c$ agrees with the results obtained by $\eta$-DDM and steady-shear measurements as well as literature values.\cite{Cheng2008}  
Small differences are only observed beyond a glycerol content of $45\;\%$. They are attributed to the uncertainty in the very low viscosity of water, $\eta(0)$, that leads to larger discrepancies in the normalized viscosities $\eta(c)/\eta(0)$ for larger absolute values of $\eta(c)$.
Nevertheless, the overall very good agreement suggests that $\eta$-DDM can successfully be applied to determine the viscosity of fluids.


\subsection{Colloidal suspensions: protein samples}

In this section, $\eta$-DDM is applied to colloidal suspensions, namely lysozyme solutions with different protein concentrations. 
The viscosity of protein solutions is determined by the molecular details of the protein as well as their interactions, which can also be tuned by the solution conditions such as $p$H, salt and cosolvent.\cite{Zhang2017, Sarangapani2013, Heinen2012} It is of fundamental and technological importance, for example for transport processes, especially in the crowded environment of living cells,\cite{Heinen2012, Ellis2001, Minton2001} for protein therapeutics\cite{Mitragotri2014, Yadav2010} and food processing.\cite{Bourne2002}

\begin{figure}[bt!]
\begin{centering}
\includegraphics[width=0.7\linewidth]{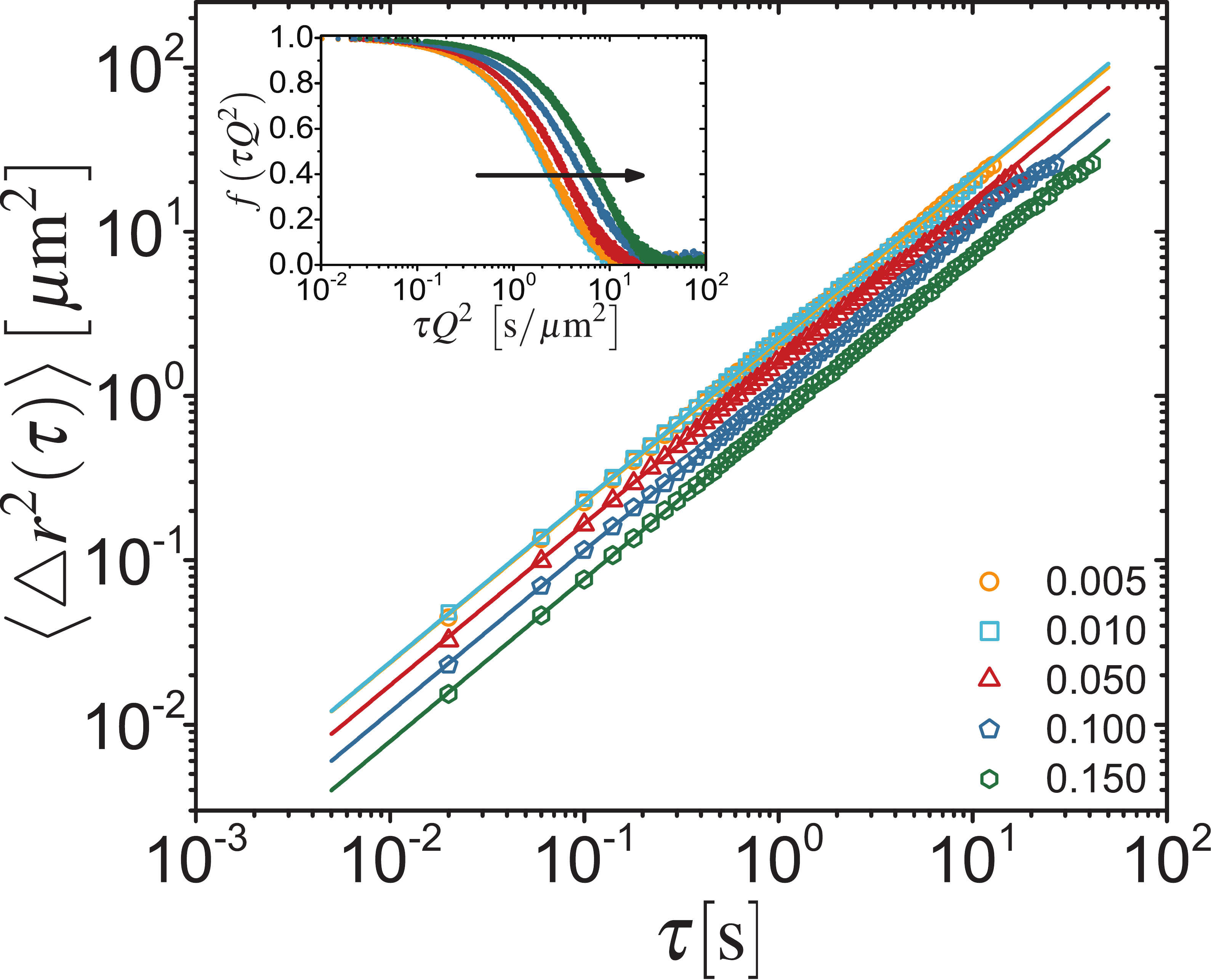}
\par\end{centering}
\protect\caption{Mean-squared displacement (MSD), $\langle \triangle r^2(\tau) \rangle$, as a function of delay time $\tau$ as determined by $\eta$-DDM for tracers coated with PEG and a diameter of $1.0\;\upmu$m in aqueous lysozyme solutions with different lysozyme concentrations (as indicated). Inset: Corresponding intermediate scattering function $f(Q,\tau)$ as a function of $\tau Q^{2}$. Note that, in this representation, $f(Q,\tau)$ for different $Q$ fall on top of each other. The arrow indicates increasing lysozyme volume fractions.}
\label{pvisco3}
\end{figure}

\begin{figure}[bt!]
\begin{centering}
\includegraphics[width=0.7\linewidth]{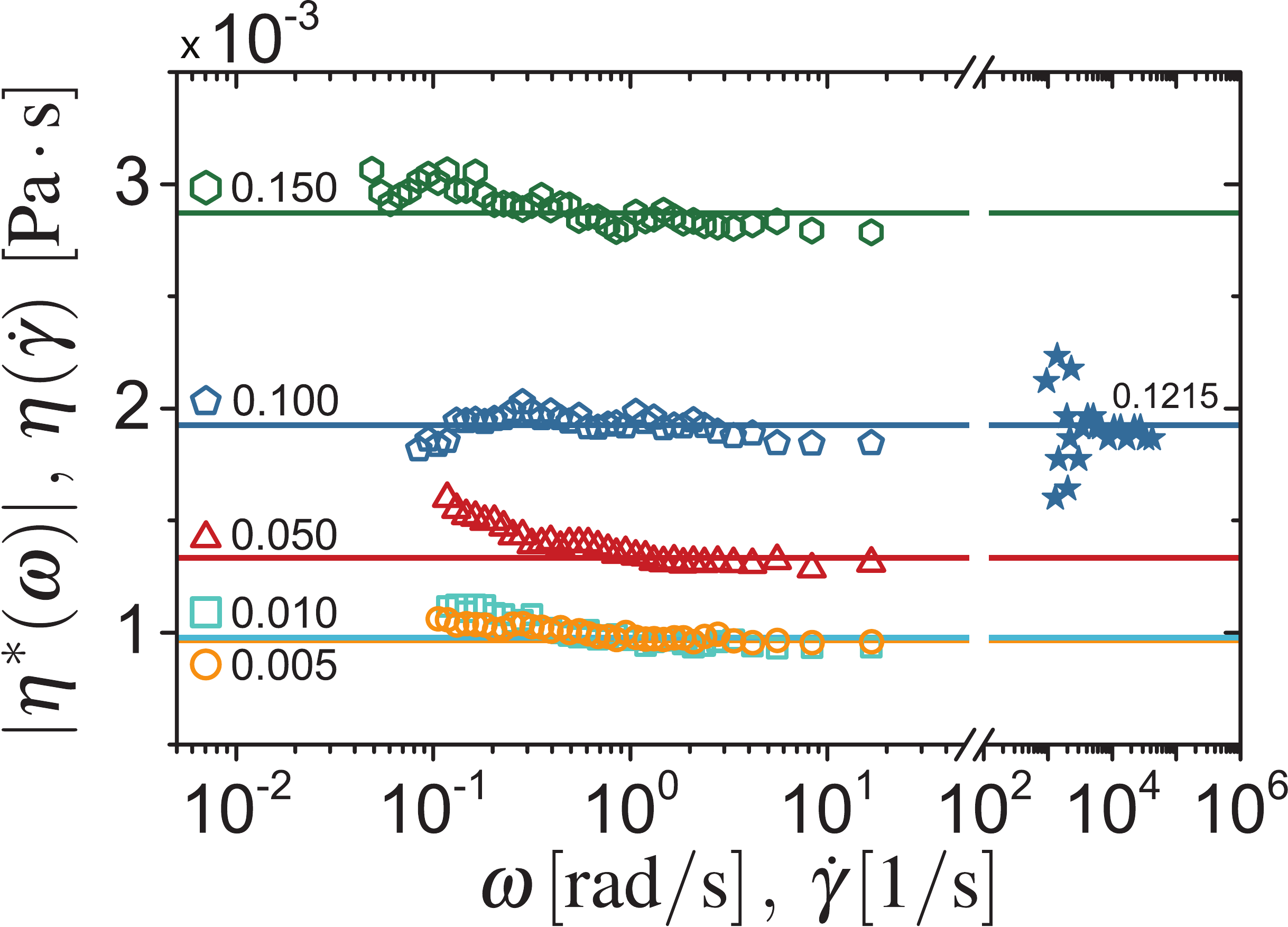}
\par\end{centering}
\protect\caption{Modulus of the complex viscosity, $|\eta^{*}(\omega)|$ and steady-shear viscosity $\eta(\dot{\gamma})$ as a function of angular frequency $\omega$ and shear rate $\dot{\gamma}$ as determined by $\eta$-DDM (open symbols) and steady-shear measurements (obtained at $T=25^\circ$~C;\cite{Godfrin2015} filled stars), respectively, of aqueous lysozyme solutions with different lysozyme volume fractions (as indicated).}
\label{pvisco4}
\end{figure}

\begin{figure}[bt!]
\begin{centering}
\includegraphics[width=0.7\linewidth]{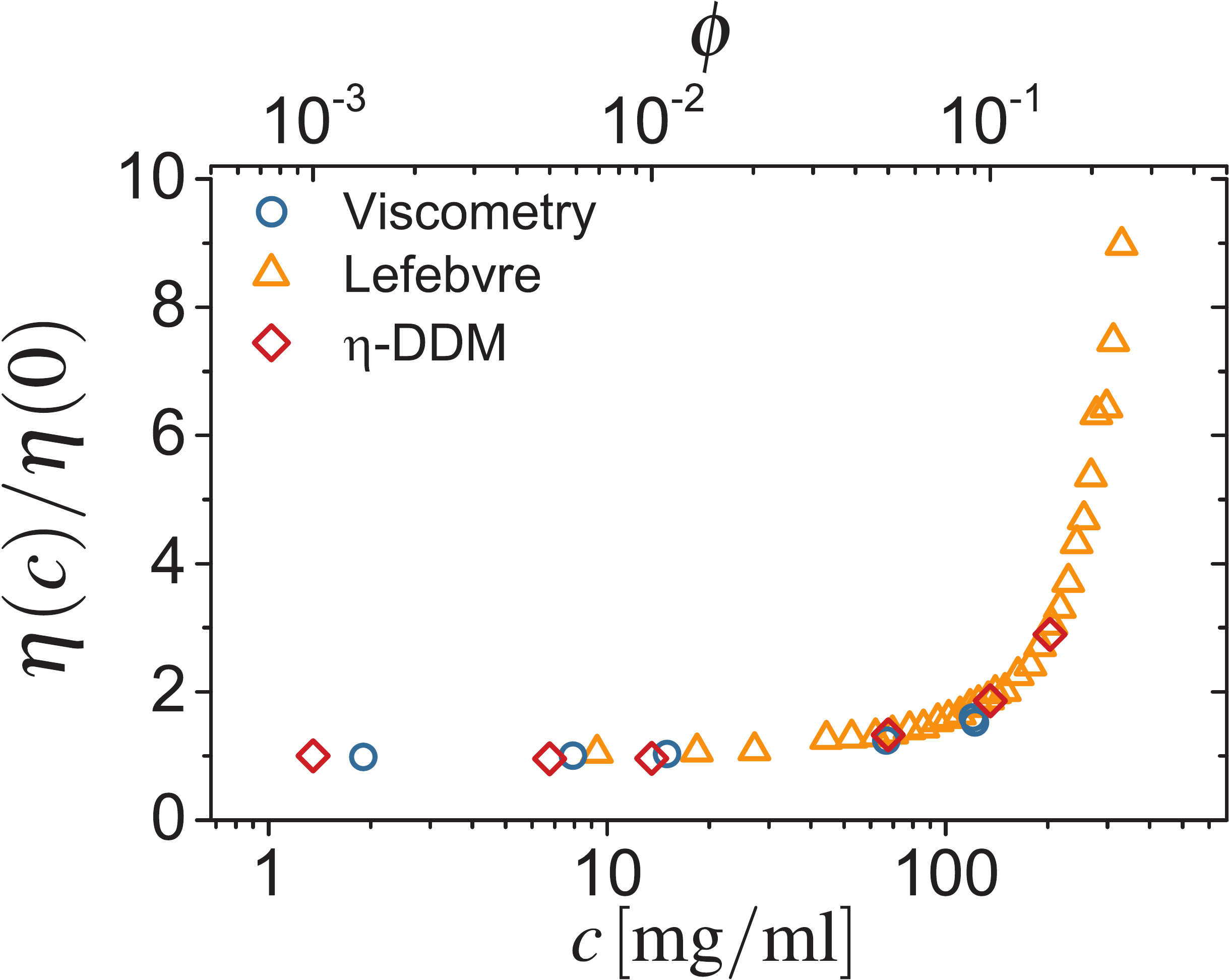}
\par\end{centering}
\protect\caption{Normalized viscosity, $\eta(c)/\eta(0)$, as a function of lysozyme concentration $c$ (bottom axis) and volume fraction $\phi$ (top axis) of lysozyme solutions as determined by $\eta$-DDM and viscometry measurements as well as literature data\cite{Lefebvre1982} (as indicated). }
\label{pvisco5}
\end{figure}

$\eta$-DDM was performed by following the same procedure as above. 
However, in order to minimize protein-particle interactions,\cite{Norde1995} polystyrene spheres coated with hydrophilic PEG 300 were used as particle tracers. The intermediate scattering function $f(Q,\tau)$ and MSD were determined (Fig.~\ref{pvisco3}). Similar to the viscous glycerol-water mixtures, the $f(Q,\tau)$ as a function of $\tau Q^2$ fall on top of each other and  both, $f(Q,\tau)$ and the MSD, shift to larger delay times $\tau$ as the protein concentration and hence the role of interactions increase. Moreover, the MSD increase linearly with the delay time $\tau$ suggesting that, within the explored parameter range, the protein solutions behave as purely viscous systems without any elastic effects. Correspondingly, the moduli of the complex viscosities $|\eta^*(\omega)|$ do not depend on the angular frequency $\omega$ (Fig.~\ref{pvisco4}). 
The magnitude of $|\eta^*(\omega)|$ is consistent with previous measurements\cite{Godfrin2015} (Fig.~\ref{pvisco4}, filled stars). Interestingly, similar qualitative and quantitative results were observed for bovine serum albumin (BSA) solutions investigated using a microfluidic rheometer 
at comparable shear rates.\cite{Sharma2011}

Contrary to the glycerol-water mixtures, we did not perform steady-shear measurements with the lysozyme solutions since they are hampered by the formation of a dense film of adsorbed protein at the air-solution interface that induces shear thinning.\cite{Sharma2011} 
However, the $\eta$-DDM results for the viscosity at different protein concentrations, $\eta(c)$, were compared with conventional viscosity measurements and literature data\cite{Lefebvre1982} (Fig.~\ref{pvisco5}). 
The agreement supports the validity of the $\eta$-DDM measurements and the proposed procedure.


\subsection{Viscoelastic fluid: aqueous poly(ethylene oxide) solution}

The rheological properties of aqueous PEO solutions are well-studied.\cite{Kulicke83, Ebagninin09, Yu94, Bahlouli13, PEO1} They are frequently used to test rheological techniques, including microrheology based on MPT,\cite{Mason1997} DWS and DLS\cite{Dasgupta2002} or, very recently, DDM.\cite{Edera2017} We investigate a PEO solution as an example for a viscoelastic fluid to illustrate the performance of $\eta$-DDM in the presence of complications arising due to viscoelasticity.

\begin{figure}[bt!]
\begin{centering}
\includegraphics[width=0.90\linewidth]{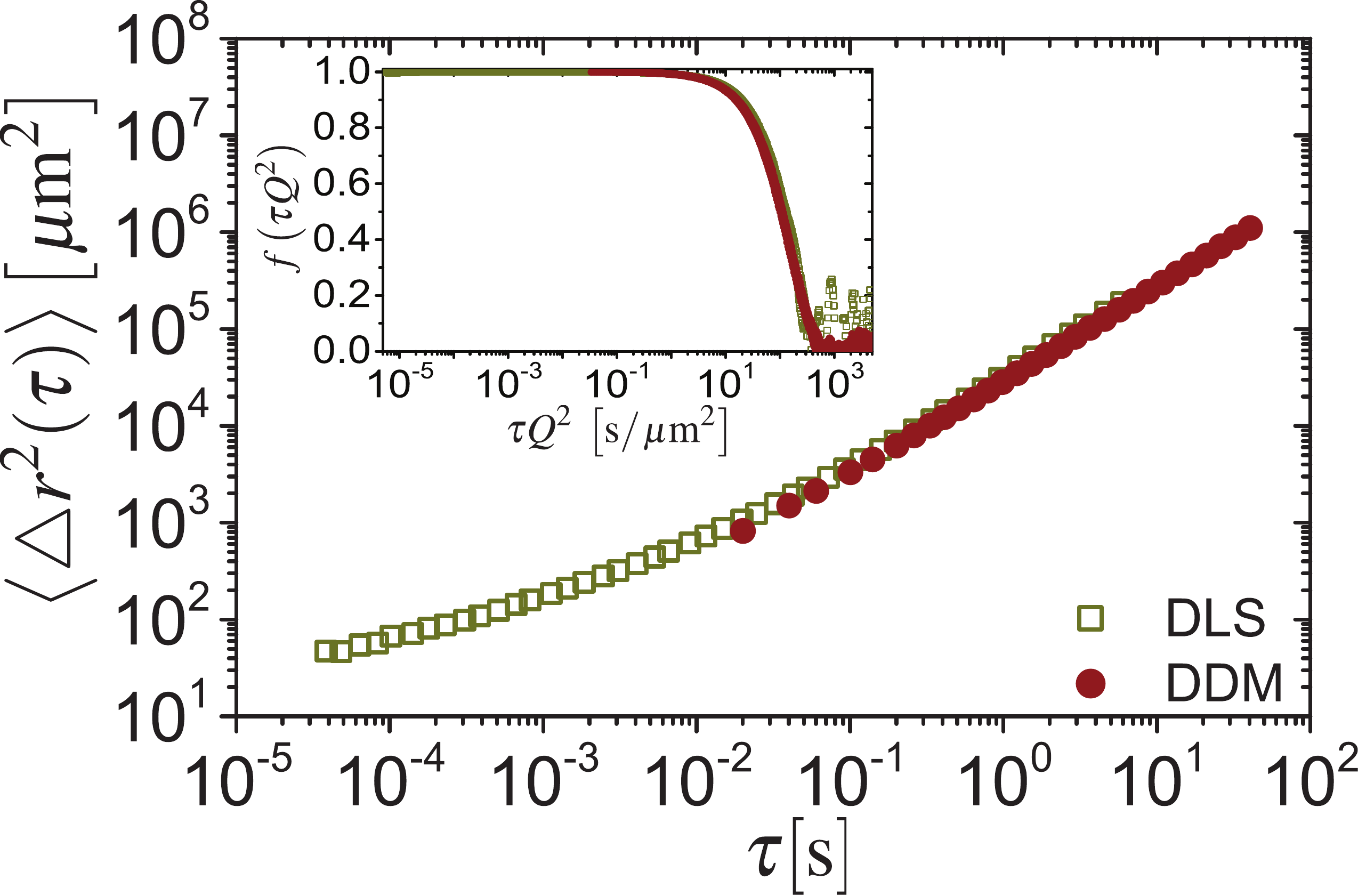}
\par\end{centering}
\protect\caption{Mean-squared displacement (MSD), $\langle \triangle r^2(\tau) \rangle$, as a function of delay time $\tau$ as determined by $\eta$-DDM (filled symbols) and DLS (open symbols) for tracers with a diameter of $330$~nm in an aqueous PEO solution. Inset: Corresponding intermediate scattering function $f(Q,\tau)$ as a function of $\tau Q^2$ for different  Q values. Note that, in this representation, $f(Q,\tau)$ for different $Q$ fall on top if each other.}
\label{pvisco1}
\end{figure}

The intermediate scattering function $f(Q,\tau)$ is determined by DDM experiments (Fig.~\ref{pvisco1}, inset). If plotted as a function of $\tau Q^2$, again, the $f(Q,\tau)$ for different $Q$ values follow the same dependence. Further analysis yields the MSD (Fig.~\ref{pvisco1}). A careful inspection of the MSD reveals that it increases linearly with delay time $\tau$ for large delay times, but deviates from this relation for short delay times. Throughout about the lowest decade covered in the experiment, the slope is smaller and hence the tracer dynamics is subdiffusive. To confirm this finding, DLS measurements were performed at three small angles ($20^\circ < \theta < 25^\circ$) which have some overlap with the $Q$ range covered in the DDM experiments. Due to the slow dynamics of the tracers, in particular on the large length scales corresponding to the small angles, as well as the single detector used in the DLS experiments, long measurement times (about $4$ days for one sample) were required to obtain statistically reliable data. The intermediate scattering function and subsequently the MSD were extracted (Eqs.~\ref{G2}, \ref{Siegert}, Fig.~\ref{pvisco1}). Since smaller delay times are accessible in DLS experiments, the nonlinear dependence is more apparent. This is consistent with previous results obtained by DDM using polymer with a higher molar mass\cite{Edera2017} and by DWS using a similar polymer but a higher concentration.\cite{Dasgupta2002} This supports the results of the DDM experiments.

\begin{figure}[bt!]
\begin{centering}
\includegraphics[width=0.85\linewidth]{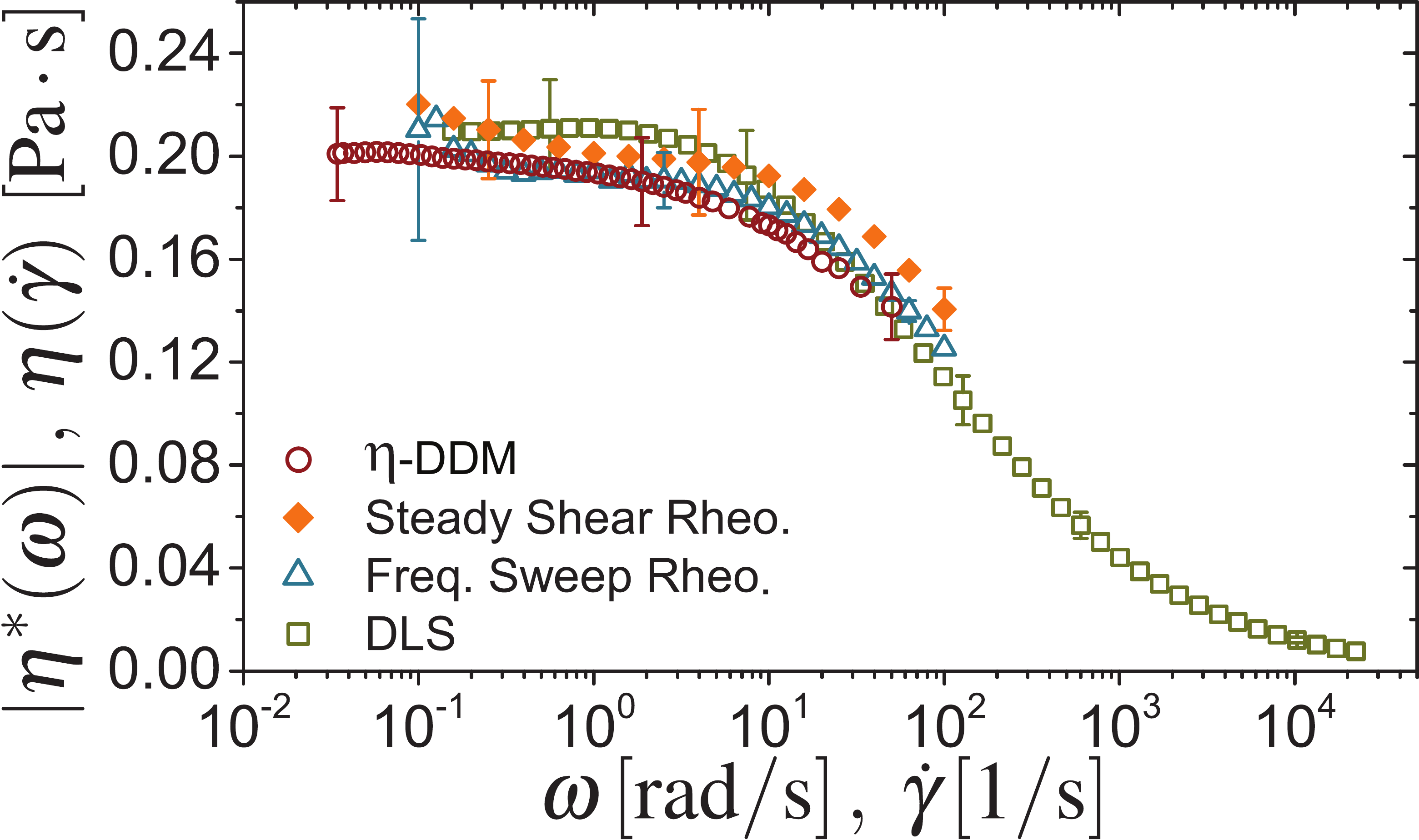}
\includegraphics[width=0.85\linewidth]{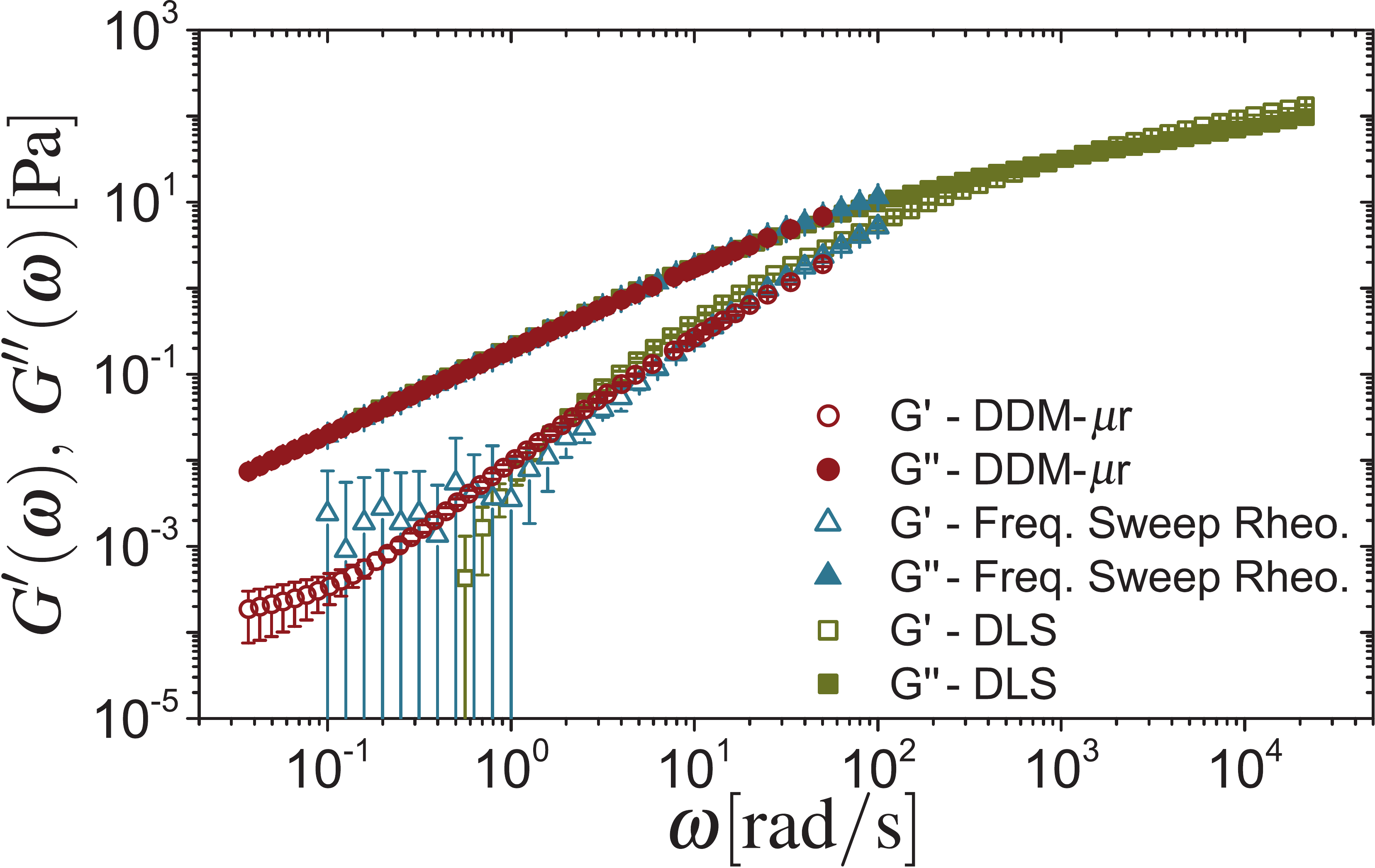}
\par\end{centering}
\protect\caption{(a) Modulus of the complex viscosity $|\eta^*(\omega)|$ and stead-shear viscosity $\eta(\dot{\gamma})$ as a function of angular frequency $\omega$ and shear rate $\dot{\gamma}$ as determined by $\eta$-DDM, SAOS, steady-shear measurements and DLS, respectively (as indicated), of an aqueous PEO solution. Note that the data are shown in a log-lin representation.
(b) Elastic, $G'(\omega)$, and loss, $G''(\omega)$, modulus extracted from DDM, SAOS and DLS measurements (as indicated) as a function of angular frequency $\omega$ of an aqueous PEO solution.
}
\label{pvisco2}
\end{figure}

Based on the MSD, the modulus of the complex viscosity $|\eta^*(\omega)|$ is calculated (Fig.~\ref{pvisco2}a). In contrast to the previous systems, it decreases with increasing $\omega$, reflecting the smaller slope of the MSD at short delay times $\tau$. To test this finding, classical rheology experiments, namely steady-shear and SAOS measurements, were performed to determine $\eta(\dot{\gamma})$ and $|\eta^*(\omega)|$, respectively (Fig.~\ref{pvisco2}a). The data of all three techniques agree, including the shear thinning behaviour at higher shear rates. This lends support to the reliability of $\eta$-DDM and confirms the applicability of the Cox-Merz rule (Eq.~\ref{coxmerz}) which previously has only been tested using classical rheometry.\cite{Yu94,Kulicke83} It also demonstrates one of the advantages of DDM. It covers relatively small $Q$ values and hence large length scales which implies large time scales or small frequencies. This is the crucial range to reliably determine the steady-shear viscosity $|\eta^*(\omega)|_{\omega\rightarrow{0}}$.

This non-trivial rheological behaviour can further be investigated based on the viscoelastic moduli, $G'(\omega)$ and $G''(\omega)$. They are also accessible by DDM (Eq.~\ref{moduli}). In the accessible frequency range, the viscous contribution, represented by the loss modulus $G''(\omega)$, is larger than the elastic contribution, $G'(\omega)$ (Fig.~\ref{pvisco2}b). In particular at low frequencies, the viscous behaviour, represented by $G''(\omega)$, dominates and reflects the approximately constant modulus of the complex viscosity and the linear increase of the MSD, i.e.~diffusive behaviour (Eqs.~\ref{Gstar1}, \ref{moduli}). In contrast, at higher frequencies both moduli converge and hence the elastic contribution becomes increasingly important, leading to the decrease in the modulus of the complex viscosity and subdiffusive dynamics with a sublinear increase of the MSD. Furthermore, the data suggest that, at frequencies $\omega$ slightly higher than accessible by DDM, the two moduli cross, i.e.~a transition from liquid to solid-like behaviour occurs. 
SAOS experiments were performed to determine the viscoelastic moduli, $G'(\omega)$ and $G''(\omega)$, in a classical rheological experiment. They agree with those from the DDM experiments (Fig.~\ref{pvisco2}b) and are also in agreement with previous results.\cite{Dasgupta2002}
In addition, DDM extends the accessible frequencies by almost a decade to lower values compared to mechanical rheology which is limited by the torque resolution of the rheometer. Furthermore, the uncertainty in the DDM data is much smaller, in particular of the inferior elastic contribution. 
This not only supports $\eta$-DDM as a method to determine the viscosity but also illustrates it versatility to comprehensively characterize the rheological behaviour of viscoelastic fluids.


\bigskip
\section{Conclusions}
\label{sec:conclusions}

DDM is increasingly used to investigate the dynamics of colloidal systems. It provides the intermediate scattering function $f(Q,\tau)$ and the mean-squared displacement (MSD) $\langle \triangle r^2(\tau) \rangle$. These parameters also allow access to rheological properties in passive microrheology experiments. In the present work, we have applied DDM to determine the viscosity. The validity of this $\eta$-DDM approach was tested for three different systems. We investigated Newtonian liquid mixtures, namely glycerol-water mixtures, and particle suspensions, that is protein solutions, as well as a viscoelastic fluid represented by an aqueous poly(ethylene oxide) solution. For all systems, a good agreement was observed between the viscosities obtained by $\eta$-DDM and conventional rheology measurements, including small amplitude oscillatory shear (SAOS) measurements, steady-shear measurements and viscometry experiments. Due to the low $Q$ range covered, DDM allows one to access frequencies up to two decades lower than in conventional rheometry which is particularly valuable for the determination of the steady-shear viscosity. Moreover, $\eta$-DDM does not require to apply an external force or displacement. Compared to conventional rheometry, in addition, only a small sample volume is required and the fact that the sample is enclosed implies that effects due to an air-sample interface are avoided.   
Thus, a conventional bright-field microscope can be used as a microliter viscometer by combining DDM, microrheology procedures and the empirical Cox-Merz rule. Hence $\eta$-DDM represents a convenient and reliable technique to determine the viscous properties of soft and biological systems.


\section*{Acknowledgements}
We thank Prof.~Roberto Cerbino (University of Milan) for very helpful discussions and Prof.~Dieter Schumacher (University D{\"u}sseldorf) for providing the capillary viscometer.
We acknowledge support from the International Helmholtz Research School of Biophysics and Soft Matter (IHRS BioSoft) and funding from the University of Granada, from the Heinrich Heine University D{\"u}sseldorf through the Strategic Research Fund as well as from the Federal Ministry for Economic Affairs and Energy (BMWi) and the German Aerospace Center (DLR) through project SoMaDy (50WM1652).





\bibliography{eta_DDM} 

\providecommand*{\mcitethebibliography}{\thebibliography}
\csname @ifundefined\endcsname{endmcitethebibliography}
{\let\endmcitethebibliography\endthebibliography}{}
\begin{mcitethebibliography}{73}
\providecommand*{\natexlab}[1]{#1}
\providecommand*{\mciteSetBstSublistMode}[1]{}
\providecommand*{\mciteSetBstMaxWidthForm}[2]{}
\providecommand*{\mciteBstWouldAddEndPuncttrue}
  {\def\EndOfBibitem{\unskip.}}
\providecommand*{\mciteBstWouldAddEndPunctfalse}
  {\let\EndOfBibitem\relax}
\providecommand*{\mciteSetBstMidEndSepPunct}[3]{}
\providecommand*{\mciteSetBstSublistLabelBeginEnd}[3]{}
\providecommand*{\EndOfBibitem}{}
\mciteSetBstSublistMode{f}
\mciteSetBstMaxWidthForm{subitem}
{(\emph{\alph{mcitesubitemcount}})}
\mciteSetBstSublistLabelBeginEnd{\mcitemaxwidthsubitemform\space}
{\relax}{\relax}

\bibitem[Walters(1975)]{Walters1975}
K.~Walters, \emph{{Rheometry}}, Wiley-VCH, 1975\relax
\mciteBstWouldAddEndPuncttrue
\mciteSetBstMidEndSepPunct{\mcitedefaultmidpunct}
{\mcitedefaultendpunct}{\mcitedefaultseppunct}\relax
\EndOfBibitem
\bibitem[Macosko(1994)]{Macosko1994}
C.~W. Macosko, \emph{{Rheology: Principles, Measurements, and Applications}},
  Wiley-VCH, 1994\relax
\mciteBstWouldAddEndPuncttrue
\mciteSetBstMidEndSepPunct{\mcitedefaultmidpunct}
{\mcitedefaultendpunct}{\mcitedefaultseppunct}\relax
\EndOfBibitem
\bibitem[Larson(1999)]{Larson1999}
R.~G. Larson, \emph{{The Structure and Rheology of Complex Fluids}}, Oxford
  University Press, 1999\relax
\mciteBstWouldAddEndPuncttrue
\mciteSetBstMidEndSepPunct{\mcitedefaultmidpunct}
{\mcitedefaultendpunct}{\mcitedefaultseppunct}\relax
\EndOfBibitem
\bibitem[Tanford(1963)]{Tanford1963}
C.~Tanford, \emph{Physical Chemistry of Macromolecules}, Wiley \& Sons: New
  York, 1963\relax
\mciteBstWouldAddEndPuncttrue
\mciteSetBstMidEndSepPunct{\mcitedefaultmidpunct}
{\mcitedefaultendpunct}{\mcitedefaultseppunct}\relax
\EndOfBibitem
\bibitem[Nicoud \emph{et~al.}(2015)Nicoud, Lattuada, Yates, and
  Morbidelli]{Nicoud2015}
L.~Nicoud, M.~Lattuada, A.~Yates and M.~Morbidelli, \emph{Soft Matter}, 2015,
  \textbf{11}, 5513--5522\relax
\mciteBstWouldAddEndPuncttrue
\mciteSetBstMidEndSepPunct{\mcitedefaultmidpunct}
{\mcitedefaultendpunct}{\mcitedefaultseppunct}\relax
\EndOfBibitem
\bibitem[Lindsay and Chaikin(1982)]{Lindsay1982}
H.~M. Lindsay and P.~M. Chaikin, \emph{J. Chem. Phys.}, 1982, \textbf{76}, 3774
  -- 3781\relax
\mciteBstWouldAddEndPuncttrue
\mciteSetBstMidEndSepPunct{\mcitedefaultmidpunct}
{\mcitedefaultendpunct}{\mcitedefaultseppunct}\relax
\EndOfBibitem
\bibitem[Sch{\"o}pe \emph{et~al.}(1998)Sch{\"o}pe, Decker, and
  Palberg]{Schoepe1998}
H.~J. Sch{\"o}pe, T.~Decker and T.~Palberg, \emph{J. Chem. Phys.}, 1998,
  \textbf{109}, 10068 -- 10074\relax
\mciteBstWouldAddEndPuncttrue
\mciteSetBstMidEndSepPunct{\mcitedefaultmidpunct}
{\mcitedefaultendpunct}{\mcitedefaultseppunct}\relax
\EndOfBibitem
\bibitem[Zhou \emph{et~al.}(2015)Zhou, Xu, Sun, and Zhu]{Zhou2015}
H.~Zhou, S.~Xu, Z.~Sun and R.~Zhu, \emph{J. Chem. Phys.}, 2015, \textbf{143},
  144903\relax
\mciteBstWouldAddEndPuncttrue
\mciteSetBstMidEndSepPunct{\mcitedefaultmidpunct}
{\mcitedefaultendpunct}{\mcitedefaultseppunct}\relax
\EndOfBibitem
\bibitem[Assael \emph{et~al.}(2014)Assael, Goodwin, Vesovic, and
  Wakeham]{Goodwin2014}
M.~J. Assael, A.~R.~H. Goodwin, V.~Vesovic and W.~A. Wakeham,
  \emph{Experimental Thermodynamics, Vol IX, Advances in Transport Properties
  of Fluids - Viscometers}, Royal Society of Chemistry, Cambridge, 2014\relax
\mciteBstWouldAddEndPuncttrue
\mciteSetBstMidEndSepPunct{\mcitedefaultmidpunct}
{\mcitedefaultendpunct}{\mcitedefaultseppunct}\relax
\EndOfBibitem
\bibitem[Hudson \emph{et~al.}(2015)Hudson, Sarangapani, Pathak, and
  Migler]{Hudson2015}
S.~D. Hudson, P.~Sarangapani, J.~A. Pathak and K.~B. Migler, \emph{J. Pharma.
  Sci.}, 2015, \textbf{104}, 678--685\relax
\mciteBstWouldAddEndPuncttrue
\mciteSetBstMidEndSepPunct{\mcitedefaultmidpunct}
{\mcitedefaultendpunct}{\mcitedefaultseppunct}\relax
\EndOfBibitem
\bibitem[Cox and Merz(1958)]{CoxMerz1958}
W.~P. Cox and E.~H. Merz, \emph{J. Polym. Sci.}, 1958, \textbf{28},
  619--622\relax
\mciteBstWouldAddEndPuncttrue
\mciteSetBstMidEndSepPunct{\mcitedefaultmidpunct}
{\mcitedefaultendpunct}{\mcitedefaultseppunct}\relax
\EndOfBibitem
\bibitem[Sharma and McKinley(2012)]{McKinley2012}
V.~Sharma and G.~H. McKinley, \emph{Rheol. Acta}, 2012, \textbf{51},
  487--495\relax
\mciteBstWouldAddEndPuncttrue
\mciteSetBstMidEndSepPunct{\mcitedefaultmidpunct}
{\mcitedefaultendpunct}{\mcitedefaultseppunct}\relax
\EndOfBibitem
\bibitem[Mason and Weitz(1995)]{Mason1995}
T.~G. Mason and D.~A. Weitz, \emph{Phys. Rev. Lett.}, 1995, \textbf{74},
  1250--1253\relax
\mciteBstWouldAddEndPuncttrue
\mciteSetBstMidEndSepPunct{\mcitedefaultmidpunct}
{\mcitedefaultendpunct}{\mcitedefaultseppunct}\relax
\EndOfBibitem
\bibitem[Waigh(2005)]{Waigh2005}
T.~A. Waigh, \emph{Rep. Prog. Phys.}, 2005, \textbf{68}, 685--742\relax
\mciteBstWouldAddEndPuncttrue
\mciteSetBstMidEndSepPunct{\mcitedefaultmidpunct}
{\mcitedefaultendpunct}{\mcitedefaultseppunct}\relax
\EndOfBibitem
\bibitem[Cicuta and Donald(2007)]{Cicuta2007}
P.~Cicuta and A.~Donald, \emph{Soft Matter}, 2007, \textbf{3}, 1449--1455\relax
\mciteBstWouldAddEndPuncttrue
\mciteSetBstMidEndSepPunct{\mcitedefaultmidpunct}
{\mcitedefaultendpunct}{\mcitedefaultseppunct}\relax
\EndOfBibitem
\bibitem[Valentine \emph{et~al.}(2004)Valentine, Perlman, Gardel, Shin,
  Matsudaira, Mitchison, and Weitz]{Valentine2004}
M.~Valentine, Z.~Perlman, M.~Gardel, J.~Shin, P.~Matsudaira, T.~Mitchison and
  D.~Weitz, \emph{Biophys. J.}, 2004, \textbf{86}, 4004--4014\relax
\mciteBstWouldAddEndPuncttrue
\mciteSetBstMidEndSepPunct{\mcitedefaultmidpunct}
{\mcitedefaultendpunct}{\mcitedefaultseppunct}\relax
\EndOfBibitem
\bibitem[Qiu \emph{et~al.}(2009)Qiu, Cosgrove, Revell, and Howell]{Qiu09}
D.~Qiu, T.~Cosgrove, P.~Revell and I.~Howell, \emph{Macromol.}, 2009,
  \textbf{42}, 547--552\relax
\mciteBstWouldAddEndPuncttrue
\mciteSetBstMidEndSepPunct{\mcitedefaultmidpunct}
{\mcitedefaultendpunct}{\mcitedefaultseppunct}\relax
\EndOfBibitem
\bibitem[Dasgupta \emph{et~al.}(2002)Dasgupta, Tee, Crocker, Frisken, and
  Weitz]{Dasgupta2002}
B.~R. Dasgupta, S.-Y. Tee, J.~C. Crocker, B.~J. Frisken and D.~A. Weitz,
  \emph{Phy. Rev. E}, 2002, \textbf{65}, 051505\relax
\mciteBstWouldAddEndPuncttrue
\mciteSetBstMidEndSepPunct{\mcitedefaultmidpunct}
{\mcitedefaultendpunct}{\mcitedefaultseppunct}\relax
\EndOfBibitem
\bibitem[Gardel \emph{et~al.}(pp.~1--49, In: Breuer K.S. (eds) {\it Microscale
  Diagnostic Techniques}, Springer, Berlin, Heidelberg, 2005)Gardel, Valentine,
  and Weitz]{Gardel05}
M.~Gardel, M.~Valentine and D.~Weitz, \emph{Microrheology}, pp.~1--49, In:
  Breuer K.S. (eds) {\it Microscale Diagnostic Techniques}, Springer, Berlin,
  Heidelberg, 2005\relax
\mciteBstWouldAddEndPuncttrue
\mciteSetBstMidEndSepPunct{\mcitedefaultmidpunct}
{\mcitedefaultendpunct}{\mcitedefaultseppunct}\relax
\EndOfBibitem
\bibitem[Squires and Mason(2010)]{Squires2010}
T.~M. Squires and T.~G. Mason, \emph{Annu. Rev. Fluid Mech.}, 2010,
  \textbf{42}, 413--438\relax
\mciteBstWouldAddEndPuncttrue
\mciteSetBstMidEndSepPunct{\mcitedefaultmidpunct}
{\mcitedefaultendpunct}{\mcitedefaultseppunct}\relax
\EndOfBibitem
\bibitem[Crocker and Grier(1996)]{Crocker1996}
J.~C. Crocker and D.~G. Grier, \emph{J. Coll. Interf. Sci.}, 1996,
  \textbf{179}, 298--310\relax
\mciteBstWouldAddEndPuncttrue
\mciteSetBstMidEndSepPunct{\mcitedefaultmidpunct}
{\mcitedefaultendpunct}{\mcitedefaultseppunct}\relax
\EndOfBibitem
\bibitem[Mason \emph{et~al.}(1997)Mason, Ganesan, van Zanten, Wirtz, and
  Kuo]{Mason1997}
T.~G. Mason, K.~Ganesan, J.~H. van Zanten, D.~Wirtz and S.~C. Kuo, \emph{Phys.
  Rev. Lett.}, 1997, \textbf{79}, 3282--3285\relax
\mciteBstWouldAddEndPuncttrue
\mciteSetBstMidEndSepPunct{\mcitedefaultmidpunct}
{\mcitedefaultendpunct}{\mcitedefaultseppunct}\relax
\EndOfBibitem
\bibitem[Kowalczyk \emph{et~al.}(2015)Kowalczyk, Oelschlaeger, and
  Willenbacher]{Kowalczyk2015}
A.~Kowalczyk, C.~Oelschlaeger and N.~Willenbacher, \emph{Meas. Sci. Technol.},
  2015, \textbf{26}, 015302\relax
\mciteBstWouldAddEndPuncttrue
\mciteSetBstMidEndSepPunct{\mcitedefaultmidpunct}
{\mcitedefaultendpunct}{\mcitedefaultseppunct}\relax
\EndOfBibitem
\bibitem[Tu and Breedveld(2005)]{Tu2005}
R.~Tu and V.~Breedveld, \emph{Phys. Rev. E}, 2005, \textbf{72}, 041914\relax
\mciteBstWouldAddEndPuncttrue
\mciteSetBstMidEndSepPunct{\mcitedefaultmidpunct}
{\mcitedefaultendpunct}{\mcitedefaultseppunct}\relax
\EndOfBibitem
\bibitem[Josephson \emph{et~al.}(2016)Josephson, Furst, and
  Galush]{Josephson2016}
L.~L. Josephson, E.~M. Furst and W.~J. Galush, \emph{J. Rheol.}, 2016,
  \textbf{60}, 531--540\relax
\mciteBstWouldAddEndPuncttrue
\mciteSetBstMidEndSepPunct{\mcitedefaultmidpunct}
{\mcitedefaultendpunct}{\mcitedefaultseppunct}\relax
\EndOfBibitem
\bibitem[Valentine \emph{et~al.}(2001)Valentine, Kaplan, Thota, Crocker,
  Gisler, Prud'homme, Beck, and Weitz]{Valentine2001}
M.~T. Valentine, P.~D. Kaplan, D.~Thota, J.~C. Crocker, T.~Gisler, R.~K.
  Prud'homme, M.~Beck and D.~A. Weitz, \emph{Phys. Rev. E}, 2001, \textbf{64},
  061506\relax
\mciteBstWouldAddEndPuncttrue
\mciteSetBstMidEndSepPunct{\mcitedefaultmidpunct}
{\mcitedefaultendpunct}{\mcitedefaultseppunct}\relax
\EndOfBibitem
\bibitem[Chen \emph{et~al.}(2010)Chen, Wen, Janmey, Crocker, and
  Yodh]{Chen2010}
D.~Chen, Q.~Wen, P.~Janmey, J.~Crocker and A.~Yodh, \emph{Annu. Rev. Condens.
  Matter Phys.}, 2010, \textbf{1}, 301--322\relax
\mciteBstWouldAddEndPuncttrue
\mciteSetBstMidEndSepPunct{\mcitedefaultmidpunct}
{\mcitedefaultendpunct}{\mcitedefaultseppunct}\relax
\EndOfBibitem
\bibitem[Giavazzi \emph{et~al.}(2009)Giavazzi, Brogioli, Trappe, Bellini, and
  Cerbino]{Giavazzi2009}
F.~Giavazzi, D.~Brogioli, V.~Trappe, T.~Bellini and R.~Cerbino, \emph{Phy. Rev.
  E}, 2009, \textbf{80}, 031403\relax
\mciteBstWouldAddEndPuncttrue
\mciteSetBstMidEndSepPunct{\mcitedefaultmidpunct}
{\mcitedefaultendpunct}{\mcitedefaultseppunct}\relax
\EndOfBibitem
\bibitem[Giavazzi and Cerbino(2014)]{Giavazzi2014}
F.~Giavazzi and R.~Cerbino, \emph{J. Opt.}, 2014, \textbf{16}, 083001\relax
\mciteBstWouldAddEndPuncttrue
\mciteSetBstMidEndSepPunct{\mcitedefaultmidpunct}
{\mcitedefaultendpunct}{\mcitedefaultseppunct}\relax
\EndOfBibitem
\bibitem[Cerbino and Trappe(2008)]{Cerbino2008}
R.~Cerbino and V.~Trappe, \emph{Phys. Rev. Lett.}, 2008, \textbf{100},
  188102\relax
\mciteBstWouldAddEndPuncttrue
\mciteSetBstMidEndSepPunct{\mcitedefaultmidpunct}
{\mcitedefaultendpunct}{\mcitedefaultseppunct}\relax
\EndOfBibitem
\bibitem[Cerbino and Cicuta(2017)]{Cerbino2017}
R.~Cerbino and P.~Cicuta, \emph{J. Chem. Phys.}, 2017, \textbf{147},
  110901\relax
\mciteBstWouldAddEndPuncttrue
\mciteSetBstMidEndSepPunct{\mcitedefaultmidpunct}
{\mcitedefaultendpunct}{\mcitedefaultseppunct}\relax
\EndOfBibitem
\bibitem[Wilson \emph{et~al.}(2011)Wilson, Martinez, Schwarz-Linek, Tailleur,
  Bryant, Pusey, and Poon]{Wilson2011}
L.~Wilson, V.~Martinez, J.~Schwarz-Linek, J.~Tailleur, G.~Bryant, P.~Pusey and
  W.~Poon, \emph{Phys. Rev. Lett.}, 2011, \textbf{106}, 018101\relax
\mciteBstWouldAddEndPuncttrue
\mciteSetBstMidEndSepPunct{\mcitedefaultmidpunct}
{\mcitedefaultendpunct}{\mcitedefaultseppunct}\relax
\EndOfBibitem
\bibitem[Martinez \emph{et~al.}(2012)Martinez, Besseling, Croze, Tailleur,
  Reufer, Schwarz-Linek, Wilson, Bees, and Poon]{Martinez2012}
V.~Martinez, R.~Besseling, O.~Croze, J.~Tailleur, M.~Reufer, J.~Schwarz-Linek,
  L.~Wilson, M.~Bees and W.~Poon, \emph{Biophys. J.}, 2012, \textbf{103},
  1637--1647\relax
\mciteBstWouldAddEndPuncttrue
\mciteSetBstMidEndSepPunct{\mcitedefaultmidpunct}
{\mcitedefaultendpunct}{\mcitedefaultseppunct}\relax
\EndOfBibitem
\bibitem[Germain \emph{et~al.}(2016)Germain, Leocmach, and Gibaud]{Germain2016}
D.~Germain, M.~Leocmach and T.~Gibaud, \emph{Am. J. Phys.}, 2016, \textbf{84},
  202--210\relax
\mciteBstWouldAddEndPuncttrue
\mciteSetBstMidEndSepPunct{\mcitedefaultmidpunct}
{\mcitedefaultendpunct}{\mcitedefaultseppunct}\relax
\EndOfBibitem
\bibitem[Bayles \emph{et~al.}(2016)Bayles, Squires, and Helgeson]{Bayles2016}
A.~Bayles, T.~Squires and M.~Helgeson, \emph{Soft Matter}, 2016, \textbf{12},
  2440--2452\relax
\mciteBstWouldAddEndPuncttrue
\mciteSetBstMidEndSepPunct{\mcitedefaultmidpunct}
{\mcitedefaultendpunct}{\mcitedefaultseppunct}\relax
\EndOfBibitem
\bibitem[Bayles \emph{et~al.}(2017)Bayles, Squires, and Helgeson]{Bayles2017}
A.~V. Bayles, T.~M. Squires and M.~E. Helgeson, \emph{Rheol. Acta}, 2017,
  \textbf{56}, 863--869\relax
\mciteBstWouldAddEndPuncttrue
\mciteSetBstMidEndSepPunct{\mcitedefaultmidpunct}
{\mcitedefaultendpunct}{\mcitedefaultseppunct}\relax
\EndOfBibitem
\bibitem[Edera \emph{et~al.}(2017)Edera, Bergamini, Trappe, Giavazzi, and
  Cerbino]{Edera2017}
P.~Edera, D.~Bergamini, V.~Trappe, F.~Giavazzi and R.~Cerbino, \emph{Phys. Rev.
  Mat.}, 2017, \textbf{1}, 073804\relax
\mciteBstWouldAddEndPuncttrue
\mciteSetBstMidEndSepPunct{\mcitedefaultmidpunct}
{\mcitedefaultendpunct}{\mcitedefaultseppunct}\relax
\EndOfBibitem
\bibitem[Grupi and Minton(2012)]{Grupi2012}
A.~Grupi and A.~Minton, \emph{Anal. Chem.}, 2012, \textbf{84},
  10732--10736\relax
\mciteBstWouldAddEndPuncttrue
\mciteSetBstMidEndSepPunct{\mcitedefaultmidpunct}
{\mcitedefaultendpunct}{\mcitedefaultseppunct}\relax
\EndOfBibitem
\bibitem[Giavazzi \emph{et~al.}(2017)Giavazzi, Edera, Lu, and
  Cerbino]{Giavazzi2017_W}
F.~Giavazzi, P.~Edera, P.~J. Lu and R.~Cerbino, \emph{Eur. Phys. J. E}, 2017,
  \textbf{40}, 97\relax
\mciteBstWouldAddEndPuncttrue
\mciteSetBstMidEndSepPunct{\mcitedefaultmidpunct}
{\mcitedefaultendpunct}{\mcitedefaultseppunct}\relax
\EndOfBibitem
\bibitem[van Megen and Underwood(1989)]{Megen89}
W.~van Megen and S.~M. Underwood, \emph{J. Chem. Phys.}, 1989, \textbf{91},
  552\relax
\mciteBstWouldAddEndPuncttrue
\mciteSetBstMidEndSepPunct{\mcitedefaultmidpunct}
{\mcitedefaultendpunct}{\mcitedefaultseppunct}\relax
\EndOfBibitem
\bibitem[Mason(2000)]{Mason2000}
T.~G. Mason, \emph{Rheol. Acta}, 2000, \textbf{39}, 371--378\relax
\mciteBstWouldAddEndPuncttrue
\mciteSetBstMidEndSepPunct{\mcitedefaultmidpunct}
{\mcitedefaultendpunct}{\mcitedefaultseppunct}\relax
\EndOfBibitem
\bibitem[Cheng(2008)]{Cheng2008}
N.-S. Cheng, \emph{Ind. Eng. Chem. Res.}, 2008, \textbf{47}, 3285--3288\relax
\mciteBstWouldAddEndPuncttrue
\mciteSetBstMidEndSepPunct{\mcitedefaultmidpunct}
{\mcitedefaultendpunct}{\mcitedefaultseppunct}\relax
\EndOfBibitem
\bibitem[Tanford and Roxby(1972)]{Tanford1972}
C.~Tanford and R.~Roxby, \emph{Biochem.}, 1972, \textbf{11}, 2192--2198\relax
\mciteBstWouldAddEndPuncttrue
\mciteSetBstMidEndSepPunct{\mcitedefaultmidpunct}
{\mcitedefaultendpunct}{\mcitedefaultseppunct}\relax
\EndOfBibitem
\bibitem[Sedgwick \emph{et~al.}(2005)Sedgwick, Kroy, Salonen, Robertson,
  Egelhaaf, and Poon]{Sedgwick2005}
H.~Sedgwick, K.~Kroy, A.~Salonen, M.~B. Robertson, S.~U. Egelhaaf and W.~C.~K.
  Poon, \emph{Eur. Phys. J. E}, 2005, \textbf{16}, 77--80\relax
\mciteBstWouldAddEndPuncttrue
\mciteSetBstMidEndSepPunct{\mcitedefaultmidpunct}
{\mcitedefaultendpunct}{\mcitedefaultseppunct}\relax
\EndOfBibitem
\bibitem[Pan \emph{et~al.}(2009)Pan, Filobelo, Pham, Galkin, Uzunova, and
  Vekilov]{Pan2009}
W.~Pan, L.~Filobelo, N.~D.~Q. Pham, O.~Galkin, V.~V. Uzunova and P.~G. Vekilov,
  \emph{Phys. Rev. Lett.}, 2009, \textbf{102}, 058101\relax
\mciteBstWouldAddEndPuncttrue
\mciteSetBstMidEndSepPunct{\mcitedefaultmidpunct}
{\mcitedefaultendpunct}{\mcitedefaultseppunct}\relax
\EndOfBibitem
\bibitem[Hansen \emph{et~al.}(2016)Hansen, Platten, Wagner, and
  Egelhaaf]{Hansen2015b}
J.~Hansen, F.~Platten, D.~Wagner and S.~U. Egelhaaf, \emph{Phys. Chem. Chem.
  Phys.}, 2016, \textbf{18}, 10270--10280\relax
\mciteBstWouldAddEndPuncttrue
\mciteSetBstMidEndSepPunct{\mcitedefaultmidpunct}
{\mcitedefaultendpunct}{\mcitedefaultseppunct}\relax
\EndOfBibitem
\bibitem[Platten \emph{et~al.}(2016)Platten, Hansen, Wagner, and
  Egelhaaf]{Platten2015c}
F.~Platten, J.~Hansen, D.~Wagner and S.~U. Egelhaaf, \emph{J. Phys. Chem.
  Lett.}, 2016, \textbf{7}, 4008--4014\relax
\mciteBstWouldAddEndPuncttrue
\mciteSetBstMidEndSepPunct{\mcitedefaultmidpunct}
{\mcitedefaultendpunct}{\mcitedefaultseppunct}\relax
\EndOfBibitem
\bibitem[Godfrin \emph{et~al.}(2015)Godfrin, Hudson, Hong, Porcar, Falus,
  Wagner, and Liu]{Godfrin2015}
P.~Godfrin, S.~Hudson, K.~Hong, L.~Porcar, P.~Falus, N.~Wagner and Y.~Liu,
  \emph{Phys. Rev. Lett.}, 2015, \textbf{115}, 228302\relax
\mciteBstWouldAddEndPuncttrue
\mciteSetBstMidEndSepPunct{\mcitedefaultmidpunct}
{\mcitedefaultendpunct}{\mcitedefaultseppunct}\relax
\EndOfBibitem
\bibitem[Platten \emph{et~al.}(2015)Platten, Valadez-P\'erez, Casta\~neda
  Priego, and Egelhaaf]{Platten2015}
F.~Platten, N.~E. Valadez-P\'erez, R.~Casta\~neda Priego and S.~U. Egelhaaf,
  \emph{J. Chem. Phys.}, 2015, \textbf{142}, 174905\relax
\mciteBstWouldAddEndPuncttrue
\mciteSetBstMidEndSepPunct{\mcitedefaultmidpunct}
{\mcitedefaultendpunct}{\mcitedefaultseppunct}\relax
\EndOfBibitem
\bibitem[Platten \emph{et~al.}(2015)Platten, Hansen, Milius, Wagner, and
  Egelhaaf]{Platten2015b}
F.~Platten, J.~Hansen, J.~Milius, D.~Wagner and S.~U. Egelhaaf, \emph{J. Phys.
  Chem. B}, 2015, \textbf{119}, 14986--14993\relax
\mciteBstWouldAddEndPuncttrue
\mciteSetBstMidEndSepPunct{\mcitedefaultmidpunct}
{\mcitedefaultendpunct}{\mcitedefaultseppunct}\relax
\EndOfBibitem
\bibitem[Jenkins and Egelhaaf(2008)]{Jenkins2008}
M.~C. Jenkins and S.~U. Egelhaaf, \emph{Adv. Coll. Interf. Sci.}, 2008,
  \textbf{136}, 65\relax
\mciteBstWouldAddEndPuncttrue
\mciteSetBstMidEndSepPunct{\mcitedefaultmidpunct}
{\mcitedefaultendpunct}{\mcitedefaultseppunct}\relax
\EndOfBibitem
\bibitem[Berne and Pecora(1976)]{Pecora}
B.~J. Berne and R.~Pecora, \emph{Dynamic Light Scattering: With Applications to
  Chemistry, Biology and Physics}, John Wiley, New York, 1976\relax
\mciteBstWouldAddEndPuncttrue
\mciteSetBstMidEndSepPunct{\mcitedefaultmidpunct}
{\mcitedefaultendpunct}{\mcitedefaultseppunct}\relax
\EndOfBibitem
\bibitem[Leffingwell \emph{et~al.}(1943)Leffingwell, Lesser, and
  Bennett]{Glycerine1}
G.~Leffingwell, M.~A. Lesser and H.~Bennett, \emph{Glycerin, Its Industrial and
  Commercial Applications}, Chemical Publishing Co., Inc., 1943\relax
\mciteBstWouldAddEndPuncttrue
\mciteSetBstMidEndSepPunct{\mcitedefaultmidpunct}
{\mcitedefaultendpunct}{\mcitedefaultseppunct}\relax
\EndOfBibitem
\bibitem[Association(1963)]{Glycerine2}
G.~P. Association, \emph{Physical Properties of Glycerine and Its Solutions},
  Glycerine Producers' Association, 1963\relax
\mciteBstWouldAddEndPuncttrue
\mciteSetBstMidEndSepPunct{\mcitedefaultmidpunct}
{\mcitedefaultendpunct}{\mcitedefaultseppunct}\relax
\EndOfBibitem
\bibitem[de~Santos~Silva and Ferreira(2012)]{Glycerine3}
M.~de~Santos~Silva and P.~C. Ferreira, \emph{Glycerol: Production, Structure,
  and Applications}, Nova Science Publishers, 2012\relax
\mciteBstWouldAddEndPuncttrue
\mciteSetBstMidEndSepPunct{\mcitedefaultmidpunct}
{\mcitedefaultendpunct}{\mcitedefaultseppunct}\relax
\EndOfBibitem
\bibitem[Segur and Oberstar(1951)]{Segur1951}
J.~B. Segur and H.~E. Oberstar, \emph{Ind. Eng. Chem.}, 1951, \textbf{43},
  2117--2120\relax
\mciteBstWouldAddEndPuncttrue
\mciteSetBstMidEndSepPunct{\mcitedefaultmidpunct}
{\mcitedefaultendpunct}{\mcitedefaultseppunct}\relax
\EndOfBibitem
\bibitem[Trejo~Gonz\'alez \emph{et~al.}(2011)Trejo~Gonz\'alez, Longinotti, and
  Corti]{Gonzalez2011}
J.~A. Trejo~Gonz\'alez, M.~P. Longinotti and H.~R. Corti, \emph{J. Chem. Eng.
  Data}, 2011, \textbf{56}, 1397--1406\relax
\mciteBstWouldAddEndPuncttrue
\mciteSetBstMidEndSepPunct{\mcitedefaultmidpunct}
{\mcitedefaultendpunct}{\mcitedefaultseppunct}\relax
\EndOfBibitem
\bibitem[Zhang and Liu(2017)]{Zhang2017}
Z.~Zhang and Y.~Liu, \emph{Curr. Opin. Chem. Eng.}, 2017, \textbf{16},
  48--55\relax
\mciteBstWouldAddEndPuncttrue
\mciteSetBstMidEndSepPunct{\mcitedefaultmidpunct}
{\mcitedefaultendpunct}{\mcitedefaultseppunct}\relax
\EndOfBibitem
\bibitem[Sarangapani \emph{et~al.}(2013)Sarangapani, Hudson, Migler, and
  Pathak]{Sarangapani2013}
P.~Sarangapani, S.~Hudson, K.~Migler and J.~Pathak, \emph{Biophys. J.}, 2013,
  \textbf{105}, 2418--2426\relax
\mciteBstWouldAddEndPuncttrue
\mciteSetBstMidEndSepPunct{\mcitedefaultmidpunct}
{\mcitedefaultendpunct}{\mcitedefaultseppunct}\relax
\EndOfBibitem
\bibitem[Heinen \emph{et~al.}(2012)Heinen, Zanini, Roosen-Runge, Fedunova,
  Zhang, Hennig, Seydel, Schweins, Sztucki, Antalik, Schreiber, and
  N\"agele]{Heinen2012}
M.~Heinen, F.~Zanini, F.~Roosen-Runge, D.~Fedunova, F.~Zhang, M.~Hennig,
  T.~Seydel, R.~Schweins, M.~Sztucki, M.~Antalik, F.~Schreiber and G.~N\"agele,
  \emph{Soft Matter}, 2012, \textbf{8}, 1404--1419\relax
\mciteBstWouldAddEndPuncttrue
\mciteSetBstMidEndSepPunct{\mcitedefaultmidpunct}
{\mcitedefaultendpunct}{\mcitedefaultseppunct}\relax
\EndOfBibitem
\bibitem[Ellis(2001)]{Ellis2001}
R.~Ellis, \emph{Trends Biochem. Sci.}, 2001, \textbf{26}, 597--604\relax
\mciteBstWouldAddEndPuncttrue
\mciteSetBstMidEndSepPunct{\mcitedefaultmidpunct}
{\mcitedefaultendpunct}{\mcitedefaultseppunct}\relax
\EndOfBibitem
\bibitem[Minton(2001)]{Minton2001}
A.~P. Minton, \emph{J. Biol. Chem.}, 2001, \textbf{276}, 10577--10580\relax
\mciteBstWouldAddEndPuncttrue
\mciteSetBstMidEndSepPunct{\mcitedefaultmidpunct}
{\mcitedefaultendpunct}{\mcitedefaultseppunct}\relax
\EndOfBibitem
\bibitem[Mitragotri \emph{et~al.}(2014)Mitragotri, Burke, and
  Langer]{Mitragotri2014}
S.~Mitragotri, P.~Burke and R.~Langer, \emph{Nat. Rev. Drug Discov.}, 2014,
  \textbf{13}, 655--672\relax
\mciteBstWouldAddEndPuncttrue
\mciteSetBstMidEndSepPunct{\mcitedefaultmidpunct}
{\mcitedefaultendpunct}{\mcitedefaultseppunct}\relax
\EndOfBibitem
\bibitem[Yadav \emph{et~al.}(2010)Yadav, Shire, and Kalonia]{Yadav2010}
S.~Yadav, S.~J. Shire and D.~S. Kalonia, \emph{J. Pharm. Sci.}, 2010,
  \textbf{99}, 4812--4829\relax
\mciteBstWouldAddEndPuncttrue
\mciteSetBstMidEndSepPunct{\mcitedefaultmidpunct}
{\mcitedefaultendpunct}{\mcitedefaultseppunct}\relax
\EndOfBibitem
\bibitem[Bourne(2002)]{Bourne2002}
M.~Bourne, \emph{Food Texture and Viscosity}, Academic Press: San Diego,
  2002\relax
\mciteBstWouldAddEndPuncttrue
\mciteSetBstMidEndSepPunct{\mcitedefaultmidpunct}
{\mcitedefaultendpunct}{\mcitedefaultseppunct}\relax
\EndOfBibitem
\bibitem[Lefebvre(1982)]{Lefebvre1982}
J.~Lefebvre, \emph{Rheol. Acta}, 1982, \textbf{21}, 620--625\relax
\mciteBstWouldAddEndPuncttrue
\mciteSetBstMidEndSepPunct{\mcitedefaultmidpunct}
{\mcitedefaultendpunct}{\mcitedefaultseppunct}\relax
\EndOfBibitem
\bibitem[Norde \emph{et~al.}(1995)Norde, Galist{\'e}o~Gonzalez, and
  Haynes]{Norde1995}
W.~Norde, F.~Galist{\'e}o~Gonzalez and C.~A. Haynes, \emph{Poly. Adv.
  Technol.}, 1995, \textbf{6}, 518--525\relax
\mciteBstWouldAddEndPuncttrue
\mciteSetBstMidEndSepPunct{\mcitedefaultmidpunct}
{\mcitedefaultendpunct}{\mcitedefaultseppunct}\relax
\EndOfBibitem
\bibitem[Sharma \emph{et~al.}(2011)Sharma, Jaishankar, Wang, and
  McKinley]{Sharma2011}
V.~Sharma, A.~Jaishankar, Y.-C. Wang and G.~H. McKinley, \emph{Soft Matter},
  2011, \textbf{7}, 5150--5160\relax
\mciteBstWouldAddEndPuncttrue
\mciteSetBstMidEndSepPunct{\mcitedefaultmidpunct}
{\mcitedefaultendpunct}{\mcitedefaultseppunct}\relax
\EndOfBibitem
\bibitem[Kulicke \emph{et~al.}(1983)Kulicke, Elsabee, Eisenbach, and
  Peuscher]{Kulicke83}
W.-M. Kulicke, M.~Elsabee, C.~D. Eisenbach and M.~Peuscher, \emph{Polym.
  Bull.}, 1983, \textbf{9}, 190--197\relax
\mciteBstWouldAddEndPuncttrue
\mciteSetBstMidEndSepPunct{\mcitedefaultmidpunct}
{\mcitedefaultendpunct}{\mcitedefaultseppunct}\relax
\EndOfBibitem
\bibitem[Ebagninin~Wilfried \emph{et~al.}(2009)Ebagninin~Wilfried, Benchabane,
  and Bekkour]{Ebagninin09}
K.~Ebagninin~Wilfried, A.~Benchabane and K.~Bekkour, \emph{J. Coll. Interf.
  Sci.}, 2009, \textbf{336}, 360--367\relax
\mciteBstWouldAddEndPuncttrue
\mciteSetBstMidEndSepPunct{\mcitedefaultmidpunct}
{\mcitedefaultendpunct}{\mcitedefaultseppunct}\relax
\EndOfBibitem
\bibitem[Yu \emph{et~al.}(1994)Yu, L.~Amidon, D.~Weiner, and H.~Goldberg]{Yu94}
D.~M. Yu, G.~L.~Amidon, N.~D.~Weiner and A.~H.~Goldberg, \emph{J. Pharm Sci.},
  1994, \textbf{83}, 1443--1449\relax
\mciteBstWouldAddEndPuncttrue
\mciteSetBstMidEndSepPunct{\mcitedefaultmidpunct}
{\mcitedefaultendpunct}{\mcitedefaultseppunct}\relax
\EndOfBibitem
\bibitem[Ilies~Bahlouli \emph{et~al.}(2013)Ilies~Bahlouli, Bekkour, Benchabane,
  Hemar, and Nemdili]{Bahlouli13}
M.~Ilies~Bahlouli, K.~Bekkour, A.~Benchabane, Y.~Hemar and A.~Nemdili,
  \emph{Appl. Rheol.}, 2013, \textbf{23}, 13435\relax
\mciteBstWouldAddEndPuncttrue
\mciteSetBstMidEndSepPunct{\mcitedefaultmidpunct}
{\mcitedefaultendpunct}{\mcitedefaultseppunct}\relax
\EndOfBibitem
\bibitem[F.E.~Bailey(1976)]{PEO1}
J.~F.E.~Bailey, \emph{Poly(ethylene Oxide)}, Elsevier Inc., 1976\relax
\mciteBstWouldAddEndPuncttrue
\mciteSetBstMidEndSepPunct{\mcitedefaultmidpunct}
{\mcitedefaultendpunct}{\mcitedefaultseppunct}\relax
\EndOfBibitem
\end{mcitethebibliography}
\bibliographystyle{rsc} 

\end{document}